\documentclass[prd,twocolumn,showpacs,groupedaddress,superscriptaddress,amsmath,amssymb]{revtex4-2}

\newcommand{\vect}[1]{\boldsymbol{#1}}

\def\nn{\nonumber}

\usepackage{color}	
\usepackage{graphicx}
\usepackage{bm}% bold math
	
\usepackage{slashed}

\usepackage[colorlinks=true, filecolor=blue, citecolor=blue,urlcolor=blue]{hyperref}

\usepackage{enumitem}

\setenumerate{label=\arabic*)}% global settings

\begin{document}
	
\abovedisplayskip = 4pt
\belowdisplayskip = 4pt
\abovedisplayshortskip = 4pt
\belowdisplayshortskip= 4pt

\title{Implication of the Weizsacker-Williams approximation\\ 
for the dark matter mediator production} 

\author{I.~V.~Voronchikhin}
\email[\textbf{e-mail}: ]{i.v.voronchikhin@gmail.com}
\affiliation{ Tomsk Polytechnic University, 634050 Tomsk, Russia}

\author{D.~V.~Kirpichnikov}
\email[\textbf{e-mail}: ]{dmbrick@gmail.com}
\affiliation{Institute for Nuclear Research, 117312 Moscow, Russia}

\begin{abstract}
The simplified connection between Standard Model~(SM) particles and light dark matter~(LDM) can be introduced via  spin-0 and spin-1 lepton-specific mediators.
Moreover, in a mediator mass range from sub-MeV to sub-GeV, fixed-target facilities such as NA64$e$, LDMX, NA64$\mu$, and M$^3$ can potentially probe such particles of the hidden sector  via  missing energy signatures that  are described by the bremsstrahlung-like process involving leptons.
We compare the Weizsaker-Williams~(WW) approximation and the exact tree-level~(ETL) approach for the bremsstrahlung-like mediator production cross section by choosing various parameters of the fixed-target experiments.
We show that the relative difference between the total cross sections calculated in the WW  and  ETL approximation varies from~$\mathcal{O}(1)~\%$ to $\mathcal{O}(10)~\%$ for a muon mode and from~$\mathcal{O}(-20)~\%$ to $\mathcal{O}(80)~\%$ for an electron mode.
We argue that  the main difference between two approaches for electron beam mode arises from peak forward region of the cross section.  We also discuss the impact of parametrization of nuclear and atomic elastic form-factors on the total cross section.
In particular, we show that the employing different form-factor parametrization can lead to a uncertainty in the cross section  at the level of~$\lesssim \mathcal{O}(10)~\%$. That study may be helpful for the  lepton fixed-target experiments that examine the bremsstrahlung-like production of the DM mediators. 
\end{abstract}

\maketitle

\section{Introduction}

Over the past few decades, astrophysical observations lead to the concept of dark matter (DM), which may manifest itself through the gravitational effects~\cite{Bergstrom:2012fi,Bertone:2016nfn}.
The evidences of the DM imply the galaxy rotation velocities, the cosmic microwave background anisotropy, the gravitational lensing, etc~\cite{Cirelli:2024ssz,Bertone:2004pz,Gelmini:2015zpa}. 
 About $\simeq 85\%$ of total mass content in our Universe is associated with the DM which cannot be explained by the Standard Model (SM)~\cite{Planck:2015fie,Planck:2018vyg}.
 Some scenarios address the DM as a solution for the anomalous magnetic moment puzzle~\cite{Aoyama:2020ynm}, the large-scale structures~\cite{Davis:1985rj}, etc.

Thermal contact between light dark matter~(LDM) and SM particles can lead to overproduction of DM particles in the early 
Universe; therefore, the relic abundance requires a depletion mechanism to yield the observed 
value~\cite{Lee:1977ua,Kolb:1985nn,Krnjaic:2015mbs}.
To avoid the LDM overproduction one can introduce the scenario with the mediator of DM (MED) that can connect LDM and SM particles via portals.
For instance, the typical scenarios and the regarding thermal target associated with  
dark boson mediators include spin-0, spin-1 and spin-2 particles such as the hidden Higgs boson~\cite{McDonald:1993ex,Burgess:2000yq,Wells:2008xg,Schabinger:2005ei,Bickendorf:2022buy,Boos:2022gtt,Sieber:2023nkq}, the dark photon~\cite{Catena:2023use,Holdom:1985ag,Izaguirre:2015yja, Essig:2010xa, Kahn:2014sra,Batell:2014mga,Izaguirre:2013uxa,Kachanovich:2021eqa,Lyubovitskij:2022hna,Gorbunov:2022dgw,Claude:2022rho,Wang:2023wrx},  and   the dark  graviton~\cite{Lee:2013bua,Kang:2020huh,Bernal:2018qlk,Folgado:2019gie,Kang:2020yul,Dutra:2019xet,Clery:2022wib,Gill:2023kyz,Wang:2019jtk,deGiorgi:2021xvm,deGiorgi:2022yha,Jodlowski:2023yne}, respectively.  

The detection of the DM particles at the accelerator-based experiments is one of a great interest of modern physics.
We note that the above-mentioned portal scenarios for LDM provide specific experimental missing energy 
signatures. Such  processes are associated with   
energy fraction of the primary beam of the charged lepton $l^\pm =(e^\pm, \mu^\pm)$ that can be carried 
away by the produced 
mediator in the bremsstrahlung-like process $l^\pm N \to l^\pm N +\mbox{MED}$, followed by invisible decay 
into pair of DM particles $\mbox{MED}\to \chi\bar{\chi}$.
The fixed-target experiments combine the advantages of high-energy particle beam of $l^\pm$ and its 
relative large intensity in order to probe LDM in the sub-GeV mass range.
In particular, as a lepton fixed-target experiments, one can exploit the existing
(NA64$e$~\cite{Gninenko:2016kpg,NA64:2016oww,NA64:2017vtt,Gninenko:2018ter,Banerjee:2019pds,Dusaev:2020gxi,Andreev:2021fzd,NA64:2022yly,NA64:2022rme,Arefyeva:2022eba,Zhevlakov:2022vio,NA64:2021acr,NA64:2021xzo} and
NA64$\mu$~\cite{Sieber:2021fue,Kirpichnikov:2021jev,NA64:2024klw}) and the projected 
( Light Dark Matter Experiment~(LDMX)~\cite{Berlin:2018bsc,Ankowski:2019mfd,Schuster:2021mlr,Akesson:2022vza} and
M$^3$~\cite{Capdevilla:2021kcf,Kahn:2018cqs}) facilities. 

One can exploit the Weizsacker-Williams (WW) 
approximation~\cite{Fermi:1925fq,vonWeizsacker:1934nji,Williams:1935dka} in order to describe 
the bremsstrahlung-like production of mediator where the charged high-energy lepton impinging on an active target.  
%The WW approach  was  developed for classical 
%electrodynamics~\cite{Jackson:1998nia}.
Next, the WW method was adapted to bremsstrahlung-like production in the cases of the one-photon exchange process~\cite{Kim:1973he,Tsai:1973py} and the photon-fusion process~\cite{Budnev:1975poe}.
In general, the WW approximation reduces the integral over momenta space and implies that energy of 
incoming particle is much greater than the particle masses~\cite{Kim:1973he}, that simplifies the 
calculation. 
The examination of the WW approximation is performed for (i) a electron mode in case of E137 
experiments~\cite{Liu:2016mqv,Liu:2017htz}, (ii) a muon mode in case of NA64$\mu$~\cite{Kirpichnikov:2021jev,Sieber:2023nkq}, and (iii) a proton bremsstrahlung ~\cite{Blumlein:2013cua,Foroughi-Abari:2021zbm,Foroughi-Abari:2024xlj,Gorbunov:2023jnx,Harland-Lang:2019zur,Gorbunov:2024vrc,Gorbunov:2024iyu}.

We emphasize that implication of the exact tree-level (ETL) method\footnote{We recall that 
ETL method implies the straightforward cross  section calculation without exploiting  any 
simplification approaches.} for a bremsstrhlung-like production of 
the hidden particles is also widely discussed in the literature for (i)~the~muon-philic 
signature~\cite{Forbes:2022bvo,Krnjaic:2024ols}, (ii)~the~lepton-flavour-violating 
process~\cite{Davoudiasl:2021mjy,Batell:2024cdl,Ponten:2024grp}, accelerator based signatures~\cite{Arefyeva:2022eba,Zhevlakov:2022vio}, and (iii)~the~hidden vector production at electron-ion 
colliders~\cite{Davoudiasl:2023pkq}.  However,  the study of both  WW and ETL methods can also be relevant for accurate 
bremsstrahlung-like signal calculation at  (i)
 International Linear Collider beam-dump experiments~\cite{Asai:2021ehn,Asai:2023dzs}, (ii)~electron beam-dump 
 experiments~\cite{Araki:2021vhy,Asai:2022zxw,Liu:2023bby,Zhevlakov:2023jzt}, (iii)~Belle II experiment~\cite{Liang:2022pul,Liang:2021kgw} 
 (that also exploits bremsstralung-like signatures), 
 and (iv)~muon beam-dump experiments~\cite{Moroi:2022qwz,Cesarotti:2023sje,Fayet:2024ddk,Rella:2022len,Radics:2023tkn}.

The WW approximation for the bremsstrahlung-like radiation of mediators has a lot of parameters that can impact the accuracy of calculations.
In particular, the accuracy of the  cross section calculation  can depend on   the parameters of the target material, the particle primary beam and its energy,  the type of the mediator, and the parametrization of the form-factor.
In addition,  we note that rate of the DM mediator production is sensitive to the maximum angle of its emission, and the 
latter can be associated with the specific design and acceptance  of the fixed-target facility  
\cite{Liu:2016mqv,Liu:2017htz}. As a result, the cut on the emission angle can  
impact the accuracy of the WW approach for some specific parameter 
space. We note however, that the implication of the mediator emission angle cut for the cross 
section  accuracy calculation was addressed in Ref.~\cite{Foroughi-Abari:2021zbm}, implying the proton-proton 
collision. 

In this paper, we study the implication of the WW approximation for calculating the cross section of different 
mediator production in various lepton fixed-target experiments. 
In particular, we continue the study of 
authors in Refs.~\cite{Gninenko:2017yus,Liu:2016mqv,Liu:2017htz,Kirpichnikov:2021jev,Sieber:2023nkq} and  calculate the production 
cross sections using the ETL and WW methods for scalar, pseudoscalar, vector and 
axial-vector mediators of SM in the case of NA64e, NA64$\mu$, LDMX, and $\mbox{M}^3$ experiments.
Also, we compare calculations of the total and differential cross sections using the ETL and WW methods for 
different maximum emission angles of the MED.
Additionally, we consider the impact of various form-factors on the cross section calculated in the WW approximation for 
the specific MED. The latter can be crucial for the bremsstrahlung-like signal estimate, since various atomic form-factor parametrizations are addressed in the literature~\cite{Bjorken:2009mm,Liu:2017htz,Davoudiasl:2023pkq}.  
    
The paper is organized as follows. In Sec.~\ref{sec:BenchModels} 
we provide models of mediators and main parameters of the considered lepton fixed-target experiments.
In Sec.~\ref{sec:BremsstrahlungLike} we describe the procedure of  cross section calculation in the case of bremsstrahlung-like production of mediator at lepton fixed-target experiments.
In Sec.~\ref{sec:crossSectionsBremssLike} 
we discuss the differential and total cross sections for different types of mediator models and lepton fixed-target experiments.
In this section we also calculate the total cross sections with corresponding missing energy 
signatures.  We conclude in Sec.~\ref{sec:Conclusion}. We collect some helpful formulas in Appendices. 

\section{Benchmark scenarios and experiments \label{sec:BenchModels}}

Let us consider now the simplified lepton-specific interaction with a light scalar, pseudosalar, vector and axial-vector mediators of DM in the following~\cite{Liu:2016mqv,Liu:2017htz}
\begin{equation}
    \mathcal{L}^{ \phi}_{\rm eff}  \supset c^{ \phi}_{ll} \,  \phi \, \overline{l} \, l, 
\end{equation}
\begin{equation}
    \mathcal{L}^{\rm P}_{\rm eff}  \supset c^{\rm P}_{ll} \,  P \, \overline{l} \gamma_5 \, l, 
\end{equation}
\begin{equation}
    \mathcal{L}^{\rm V}_{\rm eff} \supset c^{\rm V}_{ll} V'^{\mu}\overline{l}\gamma_{\mu}l,
\end{equation}
\begin{equation}
    \mathcal{L}^{\rm A}_{\rm eff} \supset c^{\rm A}_{ll} A'^{\mu}\overline{l}\gamma_5 \gamma_{\mu}l,
\end{equation}
where~$c^{\rm MED}_{ll}$ is the dimensionless coupling of lepton with 
different mediators. 
In particular, the low-energy Lagrangian for spin-0 mediator can be originated from the flavor-specific 
five-dimensional effective 
operator \cite{Batell:2017kty,Berlin:2018bsc,Forbes:2022bvo,Chen:2018vkr,Marsicano:2018vin}.

The NA64e experiment is located at the Super Proton Synchrotron~(SPS) of the European Organization for Nuclear Research~(CERN).
Protons from the SPS hitting a beryllium target with a thin lead convector produce a beam of ultra-relativistic electrons, which is scattered by the nuclei of the experiment's active thick target~\cite{NA64:2017vtt}.
The NA64$\mu$ experiment corresponds to the muon beam mode at the M2 line of the SPS in the north area of CERN SPS, which uses the detector system structure of the NA64$e$ experiment.
However, in the NA64$\mu$ experiment, an accurate reconstruction of the momentum of the incident particle is important to suppress the background processes~\cite{NA64:2024klw}.
LDMX is a planned electron beam experiment at Stanford Linear Accelerator Center~(SLAC) in which missing energy signatures are complemented by a unique technique to measure the missing momentum of electrons~\cite{Mans:2017vej,LDMX:2018cma}.
Muon Missing Momentum Experiment~(M$^3$) is a muon mode experiment that is developed at SLAC.
The M$^3$ experiment is considered  as a complementary part of the LDMX and has a similar detector base~\cite{Akesson:2022vza}.
In Table.~\ref{tab:BenchLeptonFTexp} the parameters of the considered experiments are shown.
In addition, the cut for the mediator energy ratio~$x_{\rm cut}  =E^{\rm cut}_{\rm MED}/E_l $ is used in order to 
specify  the typical missing energy cuts for the MED production cross section in the considered experiments. Note that  the total cross section of bremsstrahlung-like MED production for the specific experiment  is
    $$\sigma_{\rm tot} = \int\limits^1_{x_{\rm cut}} \! \! dx \! \! \int\limits^{\theta_{\rm max}}_0 \!\! 
    d \theta_{\rm MED} 
    \frac{d\sigma_{2\to 3}}{dx d\theta_{\rm MED}},$$
where the double differential cross section is calculated in Sec.~\ref{sec:BremsstrahlungLike} 
below for various approaches and $\theta_{\rm MED}$ is an angle between initial beam direction and momentum of the produced MED.

\section{Bremsstrahlung-like production of mediator}\label{sec:BremsstrahlungLike}

The radiation of a dark matter mediator by a lepton scattering off a heavy nucleus can be 
represented as a $2~\to~3$ process as:
\begin{equation}\label{eq:process2to3}
l^{\pm}(p)~+~N(P_i)~\rightarrow~l^{\pm}(p')~+~N(P_f)~+~\text{MED}(k),
\end{equation}
where~$p~=~(E_{l},\vect{p})$, $p'~=~(E'_{l},\vect{p'})$~-~momenta of the incoming and outgoing leptons, respectively, $k~=~(E_{\rm MED},\vect{k})$~-~momentum of the dark matter mediator, $P_i~=~(M,0)$~and~$P_f~=~(P^0_f,\vect{P}_f)$~-~momenta of the initial and final nucleus, respectively, $q~=~(q_0,\vect{q})~\equiv~P_i-P_f$~-~momentum transferred to the nucleus. We define
the virtuality of the photon in the following form:
\begin{equation*}
	t~\equiv-q^2 = -(P_i-P_f)^2 = 2 M \left(\sqrt{M^2 + P_f^2} - M\right) > 0.
\end{equation*}

The Mandelstam-like variables for the process~\eqref{eq:process2to3} can be introduced as:
\begin{equation*}
    \tilde{s} = (p' + k)^2 - m_l^2,
\quad
    \tilde{u} = (p - k)^2 - m_l^2,
\end{equation*}
\begin{equation}\label{eq:MandelstamDef2to3}
    \tilde{t} = (p - p')^2 - m_{\rm MED}^2,
\end{equation}
Also, for the introduced Mandelstam-like variables the expression~$\tilde{s}~+~\tilde{u}~+~\tilde{t}~=~-t$ is performed.
Using conditions on the mass shell for the outgoing lepton and heavy nucleus in the form~$p'~=~m_l^2$ and~$P_{f}~=~M^2$, one can obtain:
\begin{equation*}
	q_0 = -t/(2M),
	\quad
	|\vect{q}|^2 = t^2/(4M^2) + t,
\end{equation*}
\begin{equation}
	t = 2M(\sqrt{M^2 + |\vect{q}|^2} - M) \simeq |\vect{q}|^2,
\end{equation}
where we take into account that~$|\vect{q}|~\lesssim~\mathcal{O}(100)~\mbox{MeV}$ and~$M~\propto~\mathcal{O}(100)~\mbox {GeV}$, i.e. the relation~$|\vect{q}|/M~\ll~1$.

In general, the interaction of an electromagnetic field~$A_{\mu}$ and a hadron can be effectively represented as~\cite{Schwartz:2014sze}:
\begin{equation}
	\mathcal{L}^{\rm nucl}_{\rm eff}  \supset - e A_{\mu} \mathcal{J}^{\mu},
\end{equation}
where ~$\mathcal{J}^{\mu}$ is a  hadronic current~\cite{Drell:1963ej,Berestetskii:1982qgu}. 
For a heavy nucleus, one can exploit a spin-0 boson form-factor with a good 
accuracy~\cite{Beranek:2013yqa}. 
Finally, the hadronic current reads as follows in the momentum space~\cite{Rekalo:2002gv,Perdrisat:2006hj}:
\begin{equation}
	\mathcal{J}^{\mu} = F_{\rm s}(t)(P_f + P_i)^{\mu},
\end{equation}
where $F_{\rm s}(t)$~- is form-factor, which takes into account the impact of the internal structure 
through the spatial charge distribution.

The matrix element for the process~$2\to3$ can be written as:
\begin{equation}
	i \mathcal{M}_{2\to3}^{\rm MED} 
	= 
	i c^{\rm MED}_{ll} e^2 \mathcal{L}^{\mu} \left(\frac{-i\eta_{\mu \nu}}{q^2}\right) \mathcal{J}^{\nu}.
\end{equation}
For the lepton current with the radiation of spin-0 or spin-1 mediators, one can get:
\begin{multline}
	\mathcal{L}^{\mu} = 
	\overline{u}(p') 
	\left(
	\gamma^{\mu}
	\frac{\gamma_{\sigma}(p - k)^{\sigma} + m_l }{\tilde{u}}
	C_{\mathcal{L}}^{\rm MED}
	\right. + \\ + \left.
        C_{\mathcal{L}}^{\rm MED}	
	\frac{\gamma_{\sigma}(p' + k)^{\sigma} + m_l }{\tilde{s}}
	\gamma^{\mu}
	\right) 
	u(p),	
\end{multline}
where $C_{\mathcal{L}}^{\rm MED}$ is a typical term for scalar, pseudoscalar, vector and axial mediators, that take the forms, respectively:
\begin{equation*}
C_{\mathcal{L}}^{\rm S} = 1,
\;
C_{\mathcal{L}}^{\rm P} = i \gamma_{5},
\;
C_{\mathcal{L}}^{\rm V} = {\varepsilon_{\nu}}^{*}(k) \gamma^{\nu},
\;
C_{\mathcal{L}}^{\rm A} = {\varepsilon_{\nu}}^{*}(k) \gamma_{5} \gamma^{\nu},
\end{equation*}
where $\varepsilon_{\mu}(k)$ is a polarization vector for spin-1 boson.

The nuclear form-factor in the laboratory frame is associated with charge density of nucleus through the Fourier transformation in the case of heavy nuclei~\cite{Bjorken:2009mm,Kirpichnikov:2021jev,Tsai:1973py}.  
In addition, the atomic form-factor can be represented as the nuclear form-factor with screening by the Coulomb field of atomic electrons.
Indeed, the nuclear and atomic form-factors tend to~$F_{\rm nucl}(t)~\to~1$ and~$F_{\rm atom}(t)~\to~0$ as~$t~\to~0$, respectively.
Moreover, one can represent screening by the Coulomb field of atomic electrons as a convolution of the nuclear charge density with the specific screening density.

As a result, the nuclear form-factor being  multiplied by the screening term yields an
atomic form-factor~\cite{Schiff:1951zza}:
$$
F_{\rm nucl}(t) \to F_{\rm atom} (t) \equiv F_{ \rm scr} (t) F_{\rm nucl} (t), 
$$
\[ 
F_{\rm scr}(t)~=~t/(t_{\rm a}~+~t),
\]
where $\sqrt{t_{\rm a}} = 1/R_{\rm a}$ and $R_{\rm a}$ is  typical  magnitude of the atomic radius $R_{\rm a} = 111 Z^{-1/3}/m_e$. 
It is also  worth noting that a complete screening regime arises for~$t/t_{\rm a}~\ll~1$, implying that the typical atomic form-factor is sufficiently small~$F_{\rm atom}(t)~\ll~1$.
We note that the virtual photon flux for the inelastic form-factor is proportional to~$\propto~Z$; thus, this contribution can be omitted in the calculation for heavy nuclei~\cite{Bjorken:2009mm,Kirpichnikov:2021jev,Sieber:2023nkq}.

The elastic atomic Tsai-Schiff's~\cite{Tsai:1973py,Schiff:1953yzz}, Helm's~\cite{Lewin:1995rx,Dobrich:2015jyk} and exponential~\cite{Freese:1987wu} form-factors take the following forms, respectively:
\begin{equation}\label{TsaiFFdefinition11}
    F_{\rm TS}(t)  = 
    F_{\rm scr}(t)  \frac{1}{(1 + t/t_{\rm d})},
\end{equation}
\begin{align}
&    F_{\rm H}(t) 
= 
   F_{\rm scr}(t)
   \frac{3 j_1( \sqrt{t} R_{\rm H} )}{\sqrt{t} R_{\rm H}} 
   \exp{\left(- s_{\rm H}^2 t / 2 \right)},
   \label{HelmFFdefinition999}
\end{align}
\begin{equation}\label{ScreeninAtomExpFF}
    F_{\rm E}(t) =  F_{\rm scr}(t) \exp\left( - t R^2_{\rm exp} /6 \right), 
\end{equation}
where $\sqrt{t_{\rm d}} = 1/R_{\rm n}$ is the typical momentum associated with nuclear radius, $s_{\rm H} = 0.9 \; \text{fm}$ is the nuclear shell thickness, the Tsai-Schiff's, Helm's and exponential parametrizations of effective nuclear radius can be written as:
\begin{equation}
    R_{\rm n}\simeq 1/\sqrt{0.164 A^{-2/3} \text{GeV}^2},
\end{equation}
\begin{equation}
    R_{\rm exp} =  (0.91 A^{1/3} + 0.3) \text{fm},
\end{equation}
\begin{equation}
    R_{\rm H} = \sqrt{c_{\rm H}^2 + 7/3 \pi^2 a_{\rm H}^2 - 5 s_{\rm H}^2},
\end{equation}
$ a_{\rm H} = 0.52 \; \text{fm}$  and $c_{\rm H} = (1.23 A^{1/3} - 0.6) \; \text{fm}$.
One can set Helm's form-factors to zero for $t \gtrsim (4.49/R_{\rm H})^2 \equiv t_H$ due to negative values~\cite{Dobrich:2015jyk} and also use the notation $t_{\rm exp} = 1/R^2_{\rm exp}$.
Moreover, for the nucleus of interest the following inequalities hold
\[
    t_a < t_{\rm H} < t_{\rm exp} < t_{\rm d}.
\]

\begin{table}[tb]
    \begin{tabular}{lccccc}
	\hline
 \hline
	& NA64e  & LDMX & NA64$\mu$  & M$^3$ \\ \hline
        target material & Pb & Al & Pb &  W  \\ \hline
	$Z$,~\mbox{atomic number} & $82$ & $13$ & $82$ &  $74$  \\ \hline
	$A,~\mbox{g}\cdot\mbox{mole}^{-1}$  & $207$ & $27$ & $207$ &  $184$  \\ \hline
%	$\rho,~\mbox{g}\cdot\mbox{cm}^{-3}$ & $11.34$ & $2.7$ & $11.34$ &  $19.3$  \\ \hline
%	$L_T,~\mbox{cm}$ & $0.56$ & $3.56$ & $22.5$ &  $17.5$  \\ \hline
	$x_{\rm cut}=E^{\rm cut}_{\rm MED}/E_l$ & $0.5$ & $0.7$ & $0.5$ &  $0.4$  \\ \hline
	%$\eta_{\rm MED}^{\rm brem.},~\%$ & $90$  & $50$ & $90$ &  $50$  \\ \hline
	%$\eta_{\rm MED}^{\rm ann.}$,~\% & $90$ & $50$ & $90$ &  $50$  \\ \hline
	$l^\pm$,  primary beam & electron & electron & muon &  muon  \\ \hline
        $E_l$,~GeV,~\mbox{beam energy} & $100$ & $16$ & $160$ &  $15$  \\ 
        \hline
        \hline
%	$\rm LOT_{\rm current}$ & $9.37\times10^{11}$ & - & $1.98\times10^{10}$ &  -  \\ \hline
%	$\rm LOT_{\rm max}$ & $1\times10^{12}$ & $1\times10^{15}$ & $5\times10^{13}$ & $1\times10^{13}$   \\ \hline
    \end{tabular} 
    \caption{ The benchmark parameters for the total MED production cross section $l^\pm N \to l^\pm N + \mbox{MED} $ at the lepton fixed-target experiments.  Note that the $E_{\rm MED}^{\rm cut}=x_{\rm cut} E_l$ is a typical minimum missing energy threshold 
    that is associated with the specific fixed-target facility. 
    \label{tab:BenchLeptonFTexp}
	}
\end{table}

\begin{figure*}[ht!]
	\center{\includegraphics[scale=1.0]{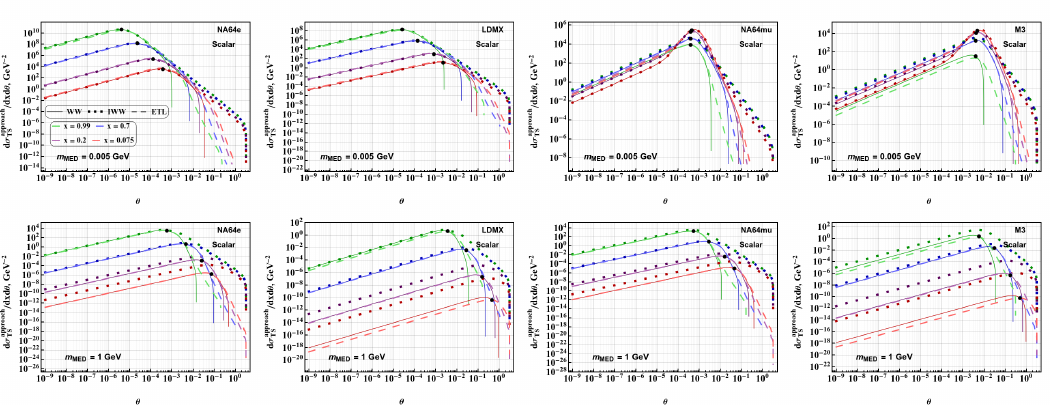}}
	\caption{ 
        The double-differential cross section with respect to the emission angle of the scalar mediator~$\theta\equiv \theta_\phi$ and its fraction of energy~$x$ as a function of~$\theta$. 
        Each color in the figures corresponds to a fixed value of fraction of energy~$x$.
        The solid, dotted and dashed lines correspond to the calculations which exploit the WW, IWW and ETL approaches, respectively.
        The typical angle of mediator radiation~\eqref{eq:TypeAngleMEDfor2D} for the each fixed value of fraction of energy~$x$ in the figures are shown by a black point.
	}\label{fig:DsDxDthetaWWandETLScalarMED}
\end{figure*}

As shown in Appendix~\ref{sec:CrossSectionBremssETL}, the ETL double-differential cross section   for the bremsstrahlung-like production of the mediator takes the form~\cite{Liu:2016mqv,Liu:2017htz}:
\begin{multline}
	\frac{d \sigma_{2\to3}}{d x d\cos(\theta_{\rm MED})}
=
	%\frac{C_{\mathcal{M}}^2}{64 \pi^3}
	\frac{(c^{\rm MED}_{ll})^2 \alpha^3 Z^2}{4 \pi}
	\frac{|\vect{k}| E_p}{|\vect{P}| |\vect{k} - \vect{p}|}
\cdot \\ \cdot
	\int\limits_{t_{\rm min}}^{t_{\rm max}}
	dt
	\frac{F^2(t)}{t^2}
	\frac{1}{8 M^2}
	\int\limits_{0}^{ 2 \pi}
	\frac{d \phi}{2 \pi}
	\left|\mathcal{A}_{2\to3}^{\rm MED} \right|^2,
\end{multline}
where~$t_{\rm max}$, $t_{\rm min}$~ are the maximum and minimum squares of the 4-momentum transferred to the 
nucleus~\eqref{eq:LimitstVirtualBremssETL}, $x~=~E_{\rm MED}/E_{l}$~ is the fraction of the total energy of the 
mediator with respect to the energy of original lepton, and
$|\mathcal{A}_{2\rightarrow 3}^{\rm MED}|^2$~ is mediator production  amplitude squared. 
In addition, the explicit forms of the squared amplitudes for the mediators are
given in~\eqref{eq:Ampl2to3ScalarMED}~-~\eqref{eq:Ampl2to3AxialVectorMED}.

Also, the double-differential cross section in the WW approximation takes the following form~\cite{Kim:1973he,Tsai:1973py}:
\begin{align}
&     \left.\frac{d \sigma ( p + P_i \rightarrow  p' + P_f + k )}{d(pk)d(kP_{i})}\right|_{WW} = 
\nn  \\   
&    
=    \frac{\alpha \chi}{\pi (p'P_i)}
    \left. \frac{d \sigma ( p + q \rightarrow  k+ p' )}{d(pk)} \right|_{ t = t_{\rm min} },
    \label{DoubleDiffWW1}  
\end{align}
where~$\alpha~=~e^2/(4\pi)~\simeq~1/137$ is the fine structure constant, and the flux of virtual photon~$\chi$ from the nucleus is expressed through the elastic form-factor~$F(t)$ as: 
\begin{equation}
\chi = 
   Z^2 \int\limits^{t_{\rm max}}_{t_{\rm min}} \frac{t - t_{\rm min}}{t^2} F^2(t) dt,
   \label{ChiDefininition1}
\end{equation}
where the typical upper limit is chosen to be 
\begin{equation}
 t_{\rm max} \simeq m_{\rm MED}^2+m^2_l,   
 \label{tmaxDefinitionWW}
\end{equation}
according to   Refs.~\cite{Liu:2016mqv,Liu:2017htz}, and $t_{\rm min}$ is defined below 
by~Eq.~(\ref{tminDefinition1}).
Using the Jacobian of the transformation from~$(k,p)$ and~$(k,\mathcal{P}_i)$ to~$\cos(\theta_{\rm MED})$ 
and~$x~=~E_{\rm MED}/E_{l}$ variables in the case of an ultra-relativistic incident lepton  in the laboratory 
frame, one can get:
\begin{multline}\label{eq:dsWW}
	\left.\frac{d \sigma ( p + P_i \rightarrow  p' + P_f + k )}{dx d\cos(\theta_{\rm MED})}\right|_{\rm WW}
	=  
	\frac{\alpha \chi}{ \pi }
\cdot \\ \cdot
	\frac{E_l^2 x \beta_{\rm MED}}{1-x}
	\left. \frac{d \sigma ( p + q \rightarrow  k + p' )}{d(pk)} \right|_{ t = t_{\rm min} },
\end{multline}
where~$\beta_{\rm MED}~=~\sqrt{1 - m_{\rm MED}^2/(x E_l)^2}$ is the typical velocity of the mediator.

Using conditions on the mass shell in the form~$p'~=~m_l^2$ and assuming the collinearity of the vector~$\vect{q}$ with the vector~$\vect{k}~-~\vect{p}$, one can obtain the minimum value of virtuality in WW approximation~$t_{\rm min}^{\rm WW}$ as:
\begin{equation} \label{tminDefinition1}
    t_{\rm min}^{\rm WW} \simeq U^2/(4E_l^2 (1-x)^2),
\end{equation}
where  we  denote the following function
\begin{equation} 
\label{eq_vect_U}
    U \! \equiv \! m_l^2 \! - \! u_2 \! \simeq \!
    E_l^2 \theta_{\rm MED}^2 x  +  m_{\rm MED}^2 (1 \! - \! x)/x  +  m_l^2 x
\!>\! 
    0.
\end{equation}

The Mandelstam variables can be expressed through the  both  $x$ and $\theta_{\rm MED}$ variables: 
\begin{equation} \label{eq_u2_U}
    u_2 = (p-k)^2 =  m_{l}^2 - U \lesssim  0,
\end{equation}
\begin{equation} \label{eq_t2_U}
    t_2 = (p-p')^2 \simeq  -Ux/(1-x) + m_{\rm MED}^2 \lesssim 0,
\end{equation}
\begin{equation} \label{eq_s2_U}
    s_2 = (p'+k)^2 \simeq U/(1 - x) + m_l^2 \gtrsim  0.
\end{equation}
In Eqs.~(\ref{tminDefinition1}),  (\ref{eq_u2_U}), (\ref{eq_t2_U}), and  (\ref{eq_s2_U}), we keep only leading terms, neglecting 
the sub-leading contributions from~\cite{Bjorken:2009mm} 
\begin{equation}
\frac{m_{\rm MED}^2}{E_{\rm MED}^2}\ll   1, \,\,
\,\, \frac{m_l^2}{(E'_l)^2}\ll   1, \,\, 
\,\, \theta_{\rm MED}\ll   1, \,\, \frac{|{\bf q}|}{E'_l} \ll   1.  
\label{SubleadingTermsForWW}
\end{equation}
%Moreover, we take into account that~$\theta_{\vect{q}\vect{p}}~\simeq~\pi$  as long as   
%$\theta_{\rm MED}~\ll~1$ and ${\bf q}$ is collinear with~$\vect{k}~-~\vect{p}$, therefore this yields, $(q, k) \simeq |\vect{q}| |\vect{k}| \simeq Ux/(2(1-x))$. 

 Both the energy conservation law, $q_0~+~E_l~=~E_l'~+~E_{\rm MED}$, and the 
 condition, $q_0 / E_l~\ll~1$, imply that  $E_l~\simeq~E_{\rm MED}~+~E'_{l}$.
Thus, one can get minimum and  maximum values of energy 
factions,~$x~=~E_{\rm MED}/E_l$, in the following 
forms~$x_{\rm min}~\simeq~m_{\rm MED}/E_l$ and $x_{\rm max}~\simeq~1~-~m_l/E_l$, respectively.

In the so-called improved WW (IWW) approach~\cite{Liu:2017htz,Gninenko:2017yus} the dependence of $t_{\rm min}$ on $x$ 
 and $\theta_{\rm MED}$ in the flux derivation is omitted to simplify 
 calculations, which is important for the integration that exploits Monte-Carlo (MC)  methods
 ~\cite{Kirpichnikov:2021jev},
 such that 
\begin{equation}
t^{\rm IWW}_{\rm min} \simeq m_{\rm MED}^4/(4 E_{l}^2).
\label{tminIWWdefinition}
\end{equation}
 This means; however, that the IWW approach is less accurate~\cite{Liu:2017htz}. We 
 also note that the following inequalities hold for the wide range of parameter 
 space, 
\begin{equation}
        t_{\rm min}^{\rm WW} 
\gtrsim 
    t_{\rm min}^{\rm IWW},
    \label{TminEneq1}
\end{equation}
\begin{equation}\label{eq:tMinApproxForxMax}
    t^{\rm WW}_{\rm min}(x = x_{\rm max}) 
\gtrsim
   \text{\rm max}
\left( \frac{ m_l^2}{4},
    \frac{E_0^4}{4 m_l^2} \theta_{\rm MED}^4 
\right),
\end{equation}
which can be helpful  for the shape analysis of the differential cross section.

In the case of spin-0 and spin-1 mediators, the matrix element for the Compton-like 
process
\begin{equation}
l^{\pm}(p)~+~\gamma(q)~\rightarrow~l^{\pm}(p')~+~\mbox{MED}(k),
\end{equation}
can be written as:
\begin{equation}
i \mathcal{M}_{2\to2}^{\rm MED}
=
i c^{\rm MED}_{ll} e \mathcal{L}^{\mu} {\varepsilon_{\mu}}(q),
\end{equation}
where $\varepsilon_{\nu}(q)$ is s polarization vector for off-shell incoming photon from nucleus.
The differential cross section for process $2 \to 2$ is: 
\begin{equation}\label{8pissMM}
    \frac{d \sigma( p + q \rightarrow  k + p' )}{d (p k)}
=
    - \frac{  (c^{\rm MED}_{ll})^2 e^2
            \left|\mathcal{A}_{2\to2}^{\rm MED} \right|^2}
        {8 \pi (s_2 - m_l^2)^2} ,
\end{equation}
where $\left|\mathcal{A}_{2\to2}^{\rm MED} \right|^2$ is  the amplitude 
squared  of the  Compton-like processes that are collected 
in~\eqref{eq:Ampl2to2ScalarMED}~-~\eqref{eq:Ampl2to2AxialVectorMED}.

\section{The  bremsstrahlung-like cross sections }\label{sec:crossSectionsBremssLike}

\begin{figure*}[ht!]
	\center{\includegraphics[scale=0.115]{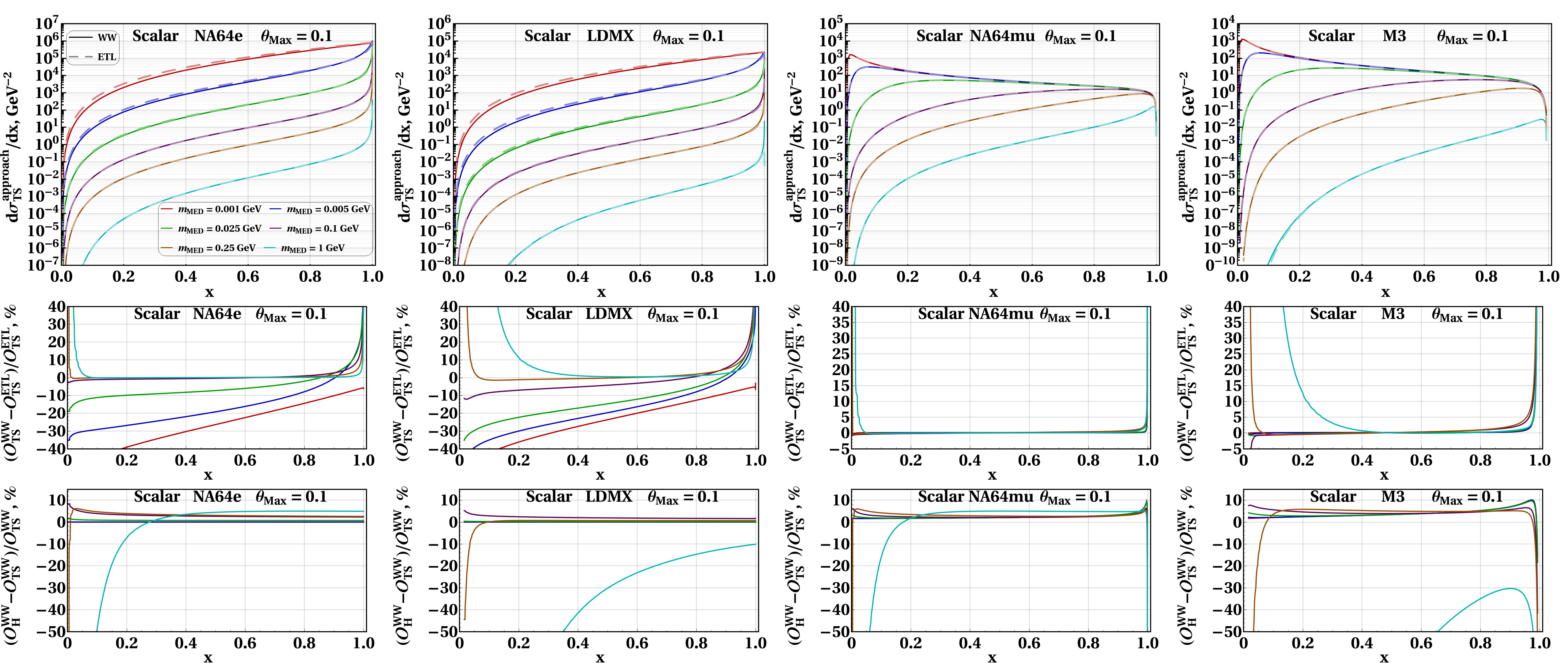}}
	\caption{ 
        The differential cross sections as a function of energy fraction x in case of scalar mediator for various experiments, such as NA64e, LDMX, NA64$\mu$, and M$^3$.
        %The differential cross sections as a function of energy fraction $x$ in the case of scalar mediator for different approximations and the corresponding relative difference, where as the benchmark  Tsai-Schiff's form-factor is chosen, also the columns correspond to NA64e, LDMX, NA64$\mu$, and M$^3$ experiments, respectively.
        We also set~$\theta_{\rm max}~=~0.1$ and~$c_{ll}^{\rm MED} = 1$.
        The curve colors correspond to fixed values of the mediator masses.
        In the upper row we show the differential cross sections for different masses, where the solid and dashed lines correspond to the WW and ETL approximations, respectively.
      In  the middle rows we show the relative difference of differential cross sections between the WW and ETL approximations.
        In the bottom row we show the relative difference between the differential cross sections calculated for Helm's~(\ref{HelmFFdefinition999})
        and  Tsai-Schiff's~(\ref{TsaiFFdefinition11})  form-factors.
	}\label{fig:DsDxWWandFFScalarMED}
\end{figure*}

In this section, we consider the cross sections in the cases of WW and ETL approximations for various benchmark  parameters. 
The total and differential cross sections are calculated by the numerical integration in the case of the  specific experimental setups. 
It is worth mentioning that the general behavior of the cross sections for scalar, pseudoscalar, vector and axial-vector mediators is similar; thus, we show for the specific cases only the differential cross section for the scalar or vector mediator.
Also, we use Tsai-Shiff's form-factor as a benchmark one 
for the  cross section in the WW and ETL approximations~\cite{Chen:2017awl,Bjorken:2009mm,Kahn:2018cqs,Kirpichnikov:2021jev}.
Moreover, in the case of the WW approximation and Tsai-Shiff's form-factor we have an 
analytical form of the differential cross section~\cite{Sieber:2023nkq}, and some auxiliary integrals for that are 
provided in Appendix~\ref{sec:AnaluticalIntegralSection}.

For a calculation of the matrix elements and a numerical integration we use the
\textsc{FeynCalc}  package~\cite{Shtabovenko:2020gxv,Shtabovenko:2016sxi} and the 
\textsc{Wolfram Mathematica} routine~\cite{Mathematica}, respectively. 
The global adaptive strategy and the Gauss–Kronrod method for numerical integration in 
\textsc{Mathematica} package are exploited.

\begin{figure*}[ht!]
	\center{\includegraphics[scale=0.115]{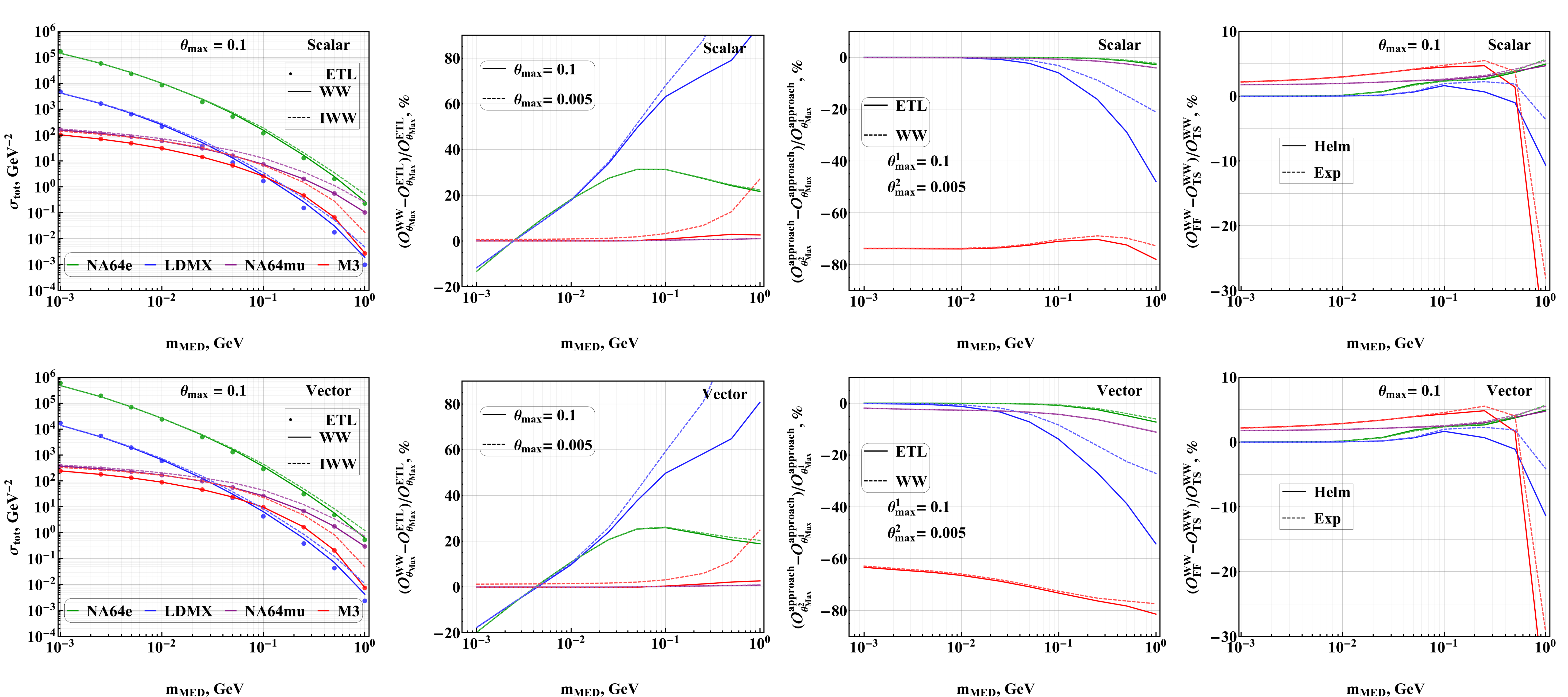}}
	\caption{  
The total cross section $\sigma_{\rm tot}$ as a function of its mass
$m_{\rm MED}$ for benchmark Tsai-Schiff’s form-factor. No energy fraction cuts are imposed on the cross sections. The first and second rows of figures 
correspond to scalar and vector mediators, respectively.   
Green, blue, purple and red lines correspond to NA64e, LDMX, NA64$\mu$ and M$^3$ experiments, respectively. In the first column the total cross sections with benchmark~$\theta_{\rm max}~=~0.1$ are shown. The solid lines, dashed lines and dots correspond to calculations for WW, IWW, and ETL approximations, respectively.
In the second column  the relative differences of total cross sections are shown for the WW and ETL approximations implying the different maximum angles of mediator emission. The solid and dotted lines correspond to the ~$\theta_{\rm max} = 0.005$ and $\theta_{\rm max} = 0.1$, respectively.
In the third column the relative differences of total cross sections are shown  for 
angles~$0.005$ and $0.1$ for the specific approximations. The solid and dashed lines 
correspond to the ETL and WW approximations, respectively.
In the fourth column  the relative differences of total cross sections are shown in the 
case of the WW approximation for different form-factors. The solid and dashed 
lines correspond to the Helm~(\ref{HelmFFdefinition999}) and 
exponential~(\ref{ScreeninAtomExpFF}) form-factors, respectively.
	}\label{fig:TCSWWandETLSVmed}
\end{figure*}

\begin{figure*}[ht!]
	\center{\includegraphics[scale=0.115]{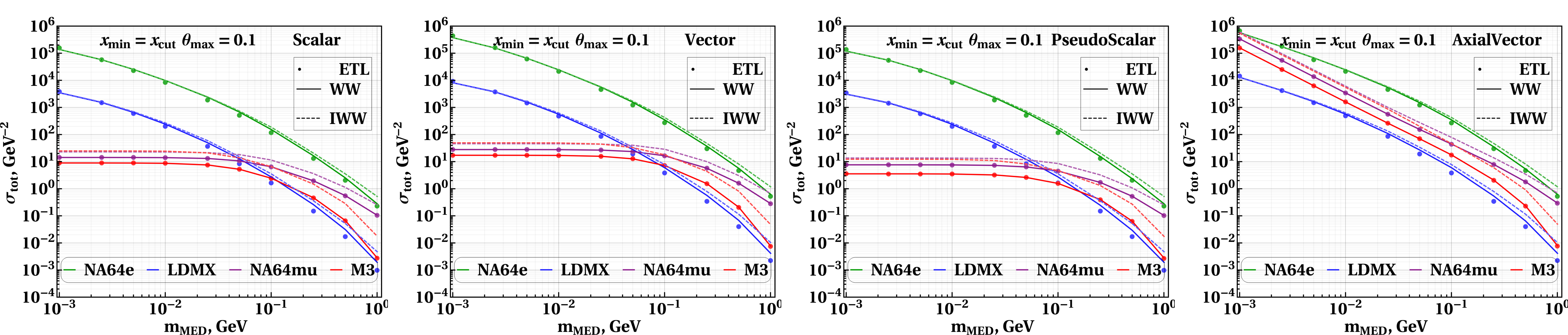}}
	\caption{  
The total cross sections~$\sigma_{\rm tot}$ with typical angle cut~$\theta_{\rm max}~=~0.1$ in the 
case of the missing energy signatures, where the benchmark  Tsai-Schiff's form-factor is chosen. 
The energy fraction cuts, $x_{\rm cut}$, for the cross sections are in 
Tab.~\ref{tab:BenchLeptonFTexp}. The plots show the cases of  scalar, vector, pseudoscalar and 
axial-vector mediators.   The
green, blue, purple and red lines correspond to NA64e, LDMX, NA64$\mu$, and M$^3$ experiments, respectively.
Solid line, dashed line and dots correspond to calculations for WW, IWW, and ETL approximations, respectively.
	}\label{fig:TCSWWandETLSPVAmed}
\end{figure*}

\subsection{The double-differential cross section}
In Fig.~\ref{fig:DsDxDthetaWWandETLScalarMED} we show the double-differential cross section as a function of MED
emission angle $\theta_{\rm MED}$ for various experiments, benchmark masses, $m_{\rm MED}$,  and 
some typical energy fractions $x$. 
The double-differential cross section has a similar dependence on the mediator radiation angle~$\theta_{\rm MED}$ with fixed~$x~=~\mbox{const}$ for each types of mediators, so that  only 
the case of scalar MED in Fig.~\ref{fig:DsDxDthetaWWandETLScalarMED} is shown.  
Also, for the fixed $x$ the double-differential cross section grows as a power low function  of 
$\theta_{\rm MED}$ to some typical angle~$\theta_{\rm typ}^{\rm 2D}$ at which it peaks and rapidly 
declines. 
Moreover, the larger benchmark value of $x$  the smaller typical angle $\theta _{\rm typ}^{\rm 2D}$ in the case of~$m_{l}~\lesssim~m_{\rm MED}$. 
In addition, the typical angle~$\theta _{\rm typ}^{\rm 2D}$ shifts slowly toward larger values of 
angle~$\theta_{\rm MED}$ for the increasing mass of the MED. 
In particular, the typical angle~$\theta _{\rm typ}^{\rm 2D}$ of the double-differential cross section lies 
in the range from~$\mathcal{O}(10^{-6})$ to~$\mathcal{O}(10^{-3})$ for~$x~\simeq 1$ in the 
 MED mass range from~$1\;\mbox{MeV}$ to~$1\;\mbox{GeV}$.
Also, for intermediate values of $x$, the typical angle~$\theta _{\rm typ}^{\rm 2D}$ takes values from~$\mathcal{O}(10^{-5})$ to~$\mathcal{O}(10^{-2})$.
For small $x\ll 1$ the typical angle~$\theta _{\rm typ}^{\rm 2D}$ ranges from from~$\mathcal{O}(10^{-4})$ to~$\mathcal{O}(10^{-1})$. 
Note that the  the region of relatively large  angle~$\theta_{\rm MED} \gtrsim \theta _{\rm typ}^{\rm 2D}$ 
the ETL cross section can be approximated by a power-law function of $\theta_{\rm MED}$; however, the WW cross section has a sharp cut-off.  
In particular, sharp cutting occurs in range from~$\mathcal{O}(10^{-2})$ to~$\mathcal{O}(10^{-1})$ in the case of 
large $x~\simeq 1$,   and in area from~$\mathcal{O}(10^{-1})$ to~$\mathcal{O}(10^{0})$ in the case of small values 
of~$x~\ll 1$. We discuss analytical derivation of the corresponding cut-off angle and $\theta^{2 \rm D}_{\rm typ}$ in Appendix~\ref{sec:TypeAngleMED}.

We note that the large values of $x$ provide the dominant contribution to the total cross section in the case of~$m_{l}~\lesssim~m_{\rm MED}$. 
As a result, one can use the typical cut-off angle~$\theta_{\rm max}~\simeq ~0.1$ for the 
total cross section calculations in order to reduce the integration time. The latter can be 
crucial for optimization  of a realistic MC simulation of Dark Sector bremsstrahlung,   implying sufficiently large statistics of leptons accumulated on 
target~\cite{Bondi:2021nfp,Oberhauser:2024ozf,NA64:2024nwj}.

%\begin{table}[tb]
%    \begin{tabular}{lccccc}\hline\hline
%    \multicolumn{5}{c}{$m_{\rm MED}~=~10^{-3}~\mbox{GeV}$} \\ 
%    \hline
%    & NA64e  & LDMX & NA64$\mu$  & M$^3$ \\ 
%    \hline
%    \hline
%	$\theta_{\rm typ}^{\rm 2D}(x_{\rm cut})$  & $8.68\cdot10^{-6}$ & $3.37\cdot10^{-5}$ &$3.81\cdot10^{-4}$ &  $4.07\cdot10^{-3}$  \\ 
%    \hline
%    $\theta_{\rm typ}^{\rm 2D}(x_{\rm max})$  & $2.95\cdot10^{-6}$ & $1.84\cdot10^{-5}$ & $3.81\cdot10^{-4}$ &  $4.07\cdot10^{-3}$  \\ 
%    \hline
%    \multicolumn{5}{c}{$m_{\rm MED}~=~1~\mbox{GeV}$} \\
%    \hline
%    & NA64e  & LDMX & NA64$\mu$  & M$^3$ \\ 
%    \hline
%	$\theta_{\rm typ}^{\rm 2D}(x_{\rm cut})$  & $8.16\cdot10^{-3}$ & $2.82\cdot10^{-2}$ & $5.12\cdot10^{-3}$ &  $7.46\cdot10^{-2}$  \\ \hline
%    $\theta_{\rm typ}^{\rm 2D}(x_{\rm max})$  & $1.33\cdot10^{-5}$ & $2.04\cdot10^{-4}$ & $3.92\cdot10^{-4}$ &  $5.21\cdot10^{-3}$  \\ \hline
%    \hline\hline\end{tabular} 
%    \caption{ 
%    \label{tab:TypAngleMEDfor2D}
%	}
%\end{table}

\subsection{The differential cross section
\label{sec:SectDiffCS}}

In Fig.~\ref{fig:DsDxWWandFFScalarMED} we show the comparison between WW and ETL differential 
cross sections for the specific parameter set.  
A description of the typical differential cross section shapes are shown in the following.
In the case of the electron mode, differential cross section has a very sharp peak in the region of the relatively large energy fraction~$x~\simeq~1$. 
Moreover, the larger $m_{\rm MED}$ the sharper peak.
In the case of the muon mode, the peak of cross section occurs near~$x \ll 1$ for small MED masses and near~$x~\simeq 1$ for large MED masses.
In addition, the peak of cross section is mitigated in the case of a muon primary beam that can be explained as follows.
%At small values of photon virtuality the cross section reaches relatively large magnitude. 
For the peak forward limit $x\simeq 1$ of electron mode, one can estimate
that $t_{\rm min}^{\rm WW} \gtrsim \mathcal{O}(m_e^2)$, thus the  cross section acquires a relatively large values in the sharp peak region (see, e.~g.~Eq.~(\ref{eq:tMinApproxForxMax}) for detail). However, for the muon 
beam a naive inequality also holds $t_{\rm min}^{\rm WW} \gtrsim \mathcal{O}(m_\mu^2)$, as a result that suppresses the muon cross section for $x\simeq 1$.
For the peak forward region 
$x \simeq 1$ the validity of the  WW approach can be broken  for the electron beam mode, as a result that can 
provide relatively large uncertainty $\mathcal{O}(80)\%$ to the total 
 cross section calculated with respect to the ETL method, as we will see below in Sec.~\ref{TotCSSubSec}.  

Next, we consider the relative difference between the differential cross sections calculated in the WW and ETL approaches for various experiments. 
\textit{NA64e}:  for the relatively large masses $m_{\rm MED} \gtrsim 100~\mbox{MeV}$,
the using of WW approximation leads to overestimation at the level of~$\mathcal{O}(10)~\%$ around~$x~\simeq 1$.
However, for intermediate and small masses of MED, an underestimation   occurs around~$\mathcal{O}(10)~\%$ in case of small values $x \ll 1$ for the WW approximation.
\textit{NA64$\mu$}: 
overestimation occurs only near the boundaries $x \ll 1$ and ~$x \simeq 1$ and 
reaches~$\mathcal{O}(10)\%$.
 \textit{LDMX} and  \textit{M$^3$}:
the behavior of the relative differences is similar to the shapes of NA64$e$ and NA64$\mu$, respectively, but the 
overestimation for the large masses starts to increase for the typical intermediate range $x\lesssim 1$.

The impact of different form-factors  on the differential cross sections in the WW 
approximation is discussed below. 
 The effects of screening and the different size description of the nucleus are associated 
 with the mediator mass at the typical scales $m_{\rm MED}\simeq (4 E_{0}^2 t_{\rm a})^{1/4}$ and~$m_{\rm MED}\simeq(4 E_{0}^2 t_{\rm d})^{1/4}$, respectively (see, e.~g.~Eq.~(\ref{TminEneq1}) for detail). Note that the screening by atomic electrons impacts on form-factors only 
for small mass of MED as~$t_{\rm min}$ reaches the typical value~$t_{\rm a}$.
In particular, the relative differences of differential cross sections with and without 
the screening are~$ \lesssim \mathcal{O}(100)~\%$ and~$ \lesssim \mathcal{O}(10)~\%$ in the cases of 
electron and muon modes implying a small mass of MED, respectively.
Similarly, the relative differences are ~$ \lesssim \mathcal{O}(1)~\%$ 
and~$\lesssim \mathcal{O}(0.1)~\%$ in the cases of electron and muon modes  for intermediate 
mass region $m_{\rm MED}\gtrsim 100~\mbox{MeV}$. Finally, for large masses of MED, $m_{\rm MED} 
\lesssim 1~\mbox{GeV}$,  one can  neglect the screening effect.

% cite, general
Let us  exploit the results of Refs.~\cite{Bjorken:2009mm,Liu:2016mqv,Liu:2017htz} for IWW differential 
cross sections to estimate the typical energy fraction for small masses of the mediator, $m_{\rm MED} \lesssim m_l$.  The authors of Refs.\cite{Bjorken:2009mm,Liu:2016mqv,Liu:2017htz} show that 
the $\theta_{\rm MED}$ can be integrated out analytically in Eq.~(\ref{eq:dsWW}) for IWW case
and  the regarding  typical differential cross section  is estimated to be
%at $x\simeq 1$
\begin{multline}
 \left(\!\frac{d \sigma_{2\to 3}}{dx} \! \right)_{\rm IWW} \!\!\!\! \propto   \text{P}_{\rm MED}(x) \! 
 \left(\! 
 m_{\rm MED}^2 \! \frac{1\!-\!x}{x}\!+\!m_l^2 x
\! \right)^{-1},   
 \label{EqIWWCSscaling}
\end{multline}
where $\text{P}_{\rm MED }(x)$ is a polynomial function of $x$ that is defined by the specific type of the mediator.

% light mediators
For sufficiently light mediators, $ m_{\rm MED} \lesssim m_l$,  one can estimate the the peak position of  the cross sections by using Eq.~(\ref{EqIWWCSscaling}).
The denominator of the Eq.~(\ref{EqIWWCSscaling}) has its minimum at $x\simeq m_{\rm MED} /m_l$.
Remarkably, that $x\simeq 
m_{\rm MED} /m_l$ is quite good approximation for the peak position of  the IWW cross section for sufficiently light mediators,  $m_{\rm MED} \lesssim m_{l}$.  
However, near~$m_{\rm MED}~\simeq~m_{l}$ the term $\text{P}_{\rm MED }(x)$ can  impact on the cross 
section peak position,  such that  $x\simeq m_{\rm MED}/m_l$   does not describe the 
position of the maximum accurately. The latter analytical calculations are rather tedious
and beyond the scope of the present paper. 

% heavy mediators
Now let us discuss typical energy fraction for sufficiently heavy mediators, $ m_{\rm MED} \gtrsim m_l$.
The kinematical constraint on $x$ implies  that $E_l \simeq E_{\rm MED} + m_l$ 
for the initial lepton transferring its entire momentum to the mediator. It  means that the typical energy fraction 
$x \lesssim  x_{\rm typ} \simeq  1- m_l/E_l$ for heavy mediators, $ m_{\rm MED} \gtrsim m_l$, at which 
the IWW cross section peaks.

% results
As results, the typical peak position of the differential 
cross section is estimated to be
\begin{equation}
 x_{\rm typ} \simeq \mbox{min} \left( 1- \frac{m_l}{E_l}, \frac{m_{\rm MED}}{m_l} \right),
 \label{eq:PeakPosX}
 \end{equation}
which holds either for  $m_{\rm MED} \lesssim m_l$ or $m_{\rm MED} \gtrsim m_l$.
Moreover, Eq.~(\ref{eq:PeakPosX}) 
also holds for both the ETL and WW cross sections, since their maxima  almost 
coincide with  the peak position of the IWW cross 
section~\cite{Liu:2016mqv,Liu:2017htz}.

We also note that the inequalities~$m_{\rm MED}/E_{\rm MED}~\ll~1$ and~$m_l/E_l'~\ll~1$ can be violated  at the boundary points
~$x~\ll~1$ and~$x~\simeq~1$, respectively. As a result, this can lead to the 
discrepancy between the ETL and WW  differential cross-section $d\sigma_{2\to3}/dx$, that is 
depicted in Fig.~\ref{fig:DsDxWWandFFScalarMED}, i.~e.~the typical magnitude of $(O_{\rm WW} - O_{\rm ETL})/O_{\rm ETL}$ can be as large as $50 \%$ for  $x\ll 1$ and $ x\simeq 1$.

%\subsection{Typical mediator energy fraction.}

\subsection{The total cross section
\label{TotCSSubSec}}

In Fig.~\ref{fig:TCSWWandETLSVmed} we show the total cross sections as a functions of $m_{\rm MED}$  
in the case of various experiments for vector and scalar MEDs without cuts.  In  Fig.~\ref{fig:TCSWWandETLSPVAmed} we 
show the total cross sections as a functions of $m_{\rm MED}$ implying the cuts  $x_{\rm cut}$ collected  
in~Table.~\ref{tab:BenchLeptonFTexp} for all types of the mediator. 
The obtained curves of the total cross section with the cuts $x_{\rm cut}$ in this section can be 
used to calculate the constraints on the corresponding coupling 
constant~\cite{Voronchikhin:2023qig} implying the missing energy signatures, 
$l^\pm N \to l^\pm N +\mbox{\rm MED}(\to \chi\bar{\chi})$. 
The latter, however, is beyond the scope of the present paper. 

Let us discuss first the impact of maximum angle of MED emission $\theta_{\rm max}$ on the validation of 
WW approach. 
The value of angle~$\theta_{\rm max}$ can be set by implying the geometry of the 
experimental detectors~\cite{Liu:2016mqv,Liu:2017htz} and the optimization of MC simulation of the MED emission 
(numerical calculations of the total and differential cross sections, say in \textsc{DMG4} package~\cite{Bondi:2021nfp,Oberhauser:2024ozf}, so that the 
setting of optimal angle  can  significantly reduce the computing time of the 
simulation).
Moreover, an optimal value of the angle~$\theta_{\rm max}$ can be considered as a parameter of interest, which 
impacts the validity  of WW approach.

However, the accuracy of calculations in the WW approximation  with different values of 
angle~$\theta_{\rm max}$ is sensitive to the choice of the  experimental parameters shown in 
Tab.~\ref{tab:BenchLeptonFTexp}.
In particular, the angle~$\theta_{\rm max}~\simeq~0.005$ was used for the analysis of the WW 
approximation~\cite{Liu:2016mqv,Liu:2017htz} due to the geometrical setup of the experiment~$\rm E137$ with  relatively long 
base-line of $\mathcal{O}(100)~\mbox{m}$. The latter is linked to the acceptance of the fixed 
target  facility implying the visible decay of $\mbox{\rm MED} \to e^+e^-$ in the fiducial 
volume and its production process $e^- N \to e^- N +\mbox{MED}$. 
Therefore, the corresponding angle~$\theta_{\rm max}~=~0.005$ is considered as a  benchmark 
one in the present article.

This is not the case for the production of the MED in the bremsstrhlung-like reaction 
$ \mu N \to \mu N +\mbox{MED}$ followed by 
its invisible decay into DM particles $\mbox{\rm MED}\to \chi \bar{\chi}$ for NA64$e$ and NA64$\mu$ facilities. 
The acceptance angle constraints $\theta \ll 1$ can be mitigated for the experiments  with short base-line 
of $\mathcal{O}(1)~\mbox{m}$, such that  the relatively large emission 
angles can be important $\theta_{\rm max} \lesssim 0.1$ for proper calculation of the
total cross section.  
In addition, the optimal value of the angle~$\theta_{\rm max}$ leads to reducing the computing time of the numerical calculations, where the contribution to the cross section in the 
range~$\theta~\gtrsim~\theta_{\rm max}$ is discarded as negligible.
In particular, the corresponding analysis was performed for the experiment 
NA64$\mu$~\cite{Kirpichnikov:2021jev,Sieber:2021fue,Sieber:2023nkq,NA64:2024klw} using the 
simulation package \textsc{DMG4}~\cite{Bondi:2021nfp,Oberhauser:2024ozf}, where the optimal value of angle was chosen to be~$\theta_{\rm max} = 0.1$. Thus, the typical value~$\theta_{\rm max}~=~0.1$ is exploited as a  second benchmark angle
in the present  paper. It is also worth noting that the ETL total cross section in the case of angles~$\theta_{\rm max}~=~0.1$ and~$\theta_{\rm max}~=~\pi$ has a relative difference at the level of~$\lesssim~\mathcal{O}(0.1)~\%$ for all experiments of interest and all mediators.

The relatively large difference between  WW and ETL total cross sections for the angles~$\theta_{\rm max}~=~0.1$ 
and~$\theta_{\rm max}~=~0.005$ arises only for the LDMX and M$^3$ experiments for 
$\mathcal{O}(100)~\mbox{MeV} \lesssim m_{\rm MED} \lesssim  \mathcal{O}(1)~\mbox{GeV}$ mass range. 
Indeed, the typical mediator radiation angle lies in the range of large angle for these experiments 
that refers to a small primary beam energy relative to the corresponding mediator mass. Also, the typical angle of MED emission can be estimated as follows $\theta \simeq m_l/E_l$. 
Moreover, in the case of the M$^3$ experiment, typical angles are  
$\simeq \mathcal{O}(10^{-2})$, which lead to a significant cross section underestimation for  
$\theta_{\rm max}~=~0.005$ if one compares it with $\theta_{\rm max}~=~0.1$.

Now let us  discuss the difference between the WW and ETL 
approximations for the typical benchmark angle~$\theta_{\rm max}~=~0.1$.
In the region of large masses, the relative difference is at the level of~$\mathcal{O}(20)~\%$ for the NA64e experiment.
However, for the LDMX experiment, the relative difference grows significantly  to the level of~$\mathcal{O}(100)~\%$ as mediator mass increases.
In the case of the NA64$\mu$ and M$^3$ experiments, the relative difference is sufficiently
small~$\lesssim \mathcal{O}(1)~\%$ for the entire range of mediator masses  and  it grows with increasing mass.
The dominant overestimation of the WW cross section is associated with a peak forward region,~$x\simeq 1$,  for  large masses $m_{\rm MED} \gtrsim 100~\mbox{MeV}$.
However, in the case of small masses $m_{\rm MED} \lesssim 100~\mbox{MeV}$, an underestimation in the WW method occurs in the region~$x~\lesssim 1$.

The relative difference of the total cross sections for different atomic form-factors is discussed below.
For the mediator of different types and
masses~$m_{\rm MED}~\lesssim~100~\mbox{MeV}$ the difference in total cross-
section is~$ \lesssim~\mathcal{O}(5)~\%$ in the case of electron and muon mode.
Also, for the NA64 experiments with high-energy leptons with~$E_{e/\mu}  \gtrsim 100~\mbox{GeV}$, the relative difference in total cross section for large 
mediator masses can be as small as~$\lesssim~\mathcal{O}(5)~\%$.
However, the relative difference for LDMX and M$^{3}$ experiments 
reaches~$\mathcal{O}(10)~\%$ for ~$m_{\rm MED} \lesssim 1~\mbox{GeV}$.

To conclude this section, let us discuss the origin of the discrepancy between ETL and WW (IWW) cross sections for the electron beam mode.
We adopt the results of Refs.~\cite{Liu:2016mqv,Liu:2017htz} that provided a link between 
 full $2 \to 3$ and  $2 \to 2$ matrix elements: 
\begin{multline}
    \frac{1}{8 M^2}
    \int \frac{d \phi}{2 \pi}
    \left|\mathcal{A}_{2\to3}^{\rm MED} \right|^2 
\simeq
    \frac{t - t_{\rm min}}{2 t_{\rm min}}
    \left|\mathcal{A}_{2\to2}^{\rm MED} \right|_{t=t_{\rm min}}^2.
    \label{eq:Link2to3and2to2}
\end{multline}
This approximate expression relies on the conditions~\eqref{SubleadingTermsForWW} and implies that
the virtual photon momenta ${\bf q}$ and $ {\bf k}~-~{\bf p}$ are highly collinear.
The explicit expressions of~$\left|\mathcal{A}_{2\to3}^{\rm MED} \right|^2$ and~$\left|\mathcal{A}_{2\to2}^{\rm MED} \right|^2$ are
given by~\eqref{eq:Ampl2to3ScalarMED}~-~\eqref{eq:Ampl2to3AxialVectorMED} and~\eqref{eq:Ampl2to2ScalarMED}~-~\eqref{eq:Ampl2to2AxialVectorMED}, respectively. Moreover,  Eq.~(\ref{eq:Link2to3and2to2}) implies that  all virtual photons  
with $t \gtrsim t_{\rm min} $ can contribute to  the mediator production.

On the other hand, in case of electron beam mode (i.~e.~for LDMX and NA64e experiments)
a main contribution 
to the single-differential cross section  is associated with the peak forward singularity  region at $x~\simeq~1$  for large masses~$m_{\rm MED} \lesssim 1~\mbox{GeV}$ of mediator.
The conditions~\eqref{SubleadingTermsForWW}
can be violated, $|{\bf q} |\lesssim E_e (1-x)$, for sufficiently large energy fraction values $x\simeq 1$ and 
relatively small energy of LDMX, $E_e\simeq 16~\mbox{GeV}$. In addition, the above-mentioned momenta collinearity 
between ${\bf q}$ and $ {\bf k} - {\bf p}$ can be also violated in the peak forward singularity region (see 
e.~g.~Refs.~\cite{Kim:1973he,Tsai:1973py,Bjorken:2009mm} for detail). 

As a result,  there is an overestimation in both IWW and WW  cross sections 
 if one compares these with ETL one. 
Say, for the LDMX experiment the regarding discrepancy can be as large as $80\%$. 
However, for the NA64e facility with larger beam energy, $E_e\simeq 100~\mbox{GeV}$,  compared to LDMX,  the error is estimated to be at the level of $20 \%$  for heavy mediator, $m_{\rm MED} \lesssim 1~\mbox{GeV}$.

\section{Conclusion\label{sec:Conclusion}}

In the present paper we have discussed in detail the calculation of the MED production cross sections  in the case of ETL and WW approaches for fixed-target experiments, such as NA64$e$, LDMX, NA64$\mu$ and M$^3$.
Scalar, pseudoscalar, vector, and axial-vector types of particle are chosen as hidden sector MEDs, that are produced in the bremsstrahlung-like reaction $l^\pm N \to l^\pm N + \mbox{MED}$.
Our study shows that the mass (electron or muon) of the incident beam particles and their energy ($E_{e/\mu} \simeq 15~\mbox{GeV}$ for LDMX and M$^3$ or $E_{e/\mu}  \gtrsim 100~\mbox{GeV}$ for NA64$e$ and NA64$\mu$)
in the fixed-target experiments can impact  the validity of the WW approach for the calculation of the bremsstrahlung-like total cross section.  
The relative difference of total cross sections varies from ~$\mathcal{O}(1)~\%$ to $\mathcal{O}(10)~\%$ for a muon mode and from~$\mathcal{O}(-20)~\%$ to $\mathcal{O}(80)~\%$ for an electron mode. The main difference between two approaches for electron beam mode arises from peak forward singularity of the cross section. 
In addition, we show that the exploiting of the various atomic form-factor parameterization leads to the cross section relative difference at the level  of~$\lesssim \mathcal{O}(10)~\%$.

\begin{acknowledgments} 
We would like to thank A.~Celentano,  P.~Crivelli, S.~Demidov, R.~Dusaev,  S.~Gninenko, D.~Gorbunov,  
M.~Kirsanov, N.~Krasnikov, E.~Kriukova, V.~Lyubovitskij, L.~Molina Bueno,  A.~Pukhov, A.~Shevelev, H.~Sieber, and 
A.~Zhevlakov    for very helpful discussions and  correspondences. This work was supported by the Foundation for the
Advancement of Theoretical Physics and Mathematics
BASIS (Project No.~\text{24-1-2-11-2} and No.~\text{24-1-2-11-1}).
This research was supported by the Ministry of Education and Science of the Russian Federation in part of the Science program (Project FSWW-2023-0003).
\end{acknowledgments}

\section{Data Availability}

The data that support the findings of this article are openly available~\cite{Voronchikhin2025hepds}.

\appendix

\iffalse
\begin{figure*}[ht!]
	\center{\includegraphics[scale=0.115]{4_dsdxETLandWWBremsVectorMEDThetaMax0.1.pdf}}
	\caption{ 
        .
	}\label{fig:DsDxWWandFFScalarMED}
\end{figure*}
\fi

\section{Cross section of bremsstrahlung-like process \label{sec:CrossSectionBremssETL}}

In this section we calculate the cross section in case of the mediator production by a lepton scattering off the heavy nucleus for the exact tree-level approach.
In particular, the differential cross section of the process~\eqref{eq:process2to3} takes the form:
\begin{equation}
	d \sigma_{2\to3}
=  
	\frac{\overline{\left| \mathcal{M}_{2\to3}^{\rm MED} \right|^2}}{4 I}
	d \Pi,
\end{equation}
and the Lorentz invariant volume of momentum space is:
\begin{equation} 
	d \Pi
=
        \frac{
	(2 \pi)^4 \delta^{(4)}(p + q - p' - k)
        d\vect{p'}d\vect{P_f}d\vect{k}}
        {(2 \pi)^3 2E_{p'} (2 \pi)^3 2E_{P_f} (2 \pi)^3 2E_{k} },
\end{equation}
where $I =\sqrt{(p,P_i)^2 - m_l^2M^2} = |\vect{p}| M$ is Miller's invariant, which characterizes the flow of initial particles.
For integration over the final states of the process~$2\to3$, one can choose four independent variables as: the azimuthal~$\phi_{\rm q}$ angle of the vector~$\vect{q}~=~-\vect{P}_f$, the virtuality of photon~$t$, angle~$\theta_{\rm MED}~=~\angle(\vect{k},\vect{p})$ and the energy ratio of the mediator and initial particle~$x~=~k_0/p_0$~\cite{Liu:2016mqv}.

Next, to obtain a double differential cross section based on the mediator parameters, we use the following frame.
The~$Oz$ axis is parallel to the spatial part of the vector~$V~\equiv~k-p$ and the vector~$\vect{k}$ lies in the~$xOz$ plane.
One can get the following formulas:
\begin{equation}\label{eq:thetaqAsFuncQ}
	\cos(\theta_{\rm q})
=
	\left( |\vect{q}|^2 + |\vect{V}|^2 - |\vect{p'}|^2 \right) / \left( 2 |\vect{q}| |\vect{V}| \right),
%	\frac{|\vect{q}|^2 + |\vect{V}|^2 - |\vect{p'}|^2}{2 |\vect{q}| |\vect{V}|},
\end{equation}
\begin{equation}
	|\vect{V}|^2 
= 
	|\vect{p}|^2 + |\vect{k}|^2  - 2 |\vect{p}| |\vect{k}| \cos(\theta_{\rm MED}),
\end{equation}
where we exploit~$|\vect{p'}|^2~=~( q_0 - V_0 )^2 - m_l^2$ and $V_0~=~- E_l(1 - x)~<~0$.
However, by taking into account inequalities~$q_0~<~0$ and~$E_l'~>~m_l$, a kinematic constraint on the photon virtually yields 
\begin{equation}
    t \leq 2 M ( E_l(1 - x) - m_l ).
\end{equation}
By taking into account the expression~$(q~-~V)^2~=~{p'}^2$, the spatial transferred momentum~$\vect{q}$ is expressed through the polar angle~$\theta_{\rm q}$ as:
\begin{equation}
	|\vect{q}| 
=
	\frac{		|\vect{V}| \cos(\theta_{\rm q}) (\tilde{u} + 2 M d_{V_0} ) 
			+ d_{V_0}\sqrt{  D } 
		 }
		 {2 d_{V_0}^2 - 2 |\vect{V}|^2 \cos^2(\theta_{\rm q}) },
\end{equation}
where we introduce the notations:
$$D~=~4 M^2 |\vect{V}|^2 \cos^2(\theta_{\rm q}) + \tilde{u}^2  + 4 M d_{V_0} \tilde{u},$$
$$d_{V_0} = (M~-~V_0)~=~E_{l}'~+~{P_f}_0~>~0.$$

Further, the Mandelstam-like variables~$\tilde{s}$ and~$\tilde{u}$ can be expressed through the selected independent variables in the following form:
\begin{equation}
	\tilde{u} = m_{\rm MED}^2 - 2 (p, k) 
% = V_0^2 - |\vect{V}|^2 - m_l^2,
\quad
	\tilde{s} = -t + 2(p,q),
\end{equation}
where corresponding dot products are:
\begin{equation*}
    (p, k) = E_0^2 x - |\vect{p}| |\vect{k}| \cos(\theta_{\rm MED}),
\end{equation*}
\begin{equation*}
    (p,q) =  q_0 E_l -  (\vect{p}_x \vect{q}_x + \vect{p}_z \vect{q}_z).
\end{equation*}
Also, by taking into account that vector~$\vect{p}$ lies in the plane~$xOz$, the projections of momenta~$\vect{p}$ and~$\vect{q}$ take the form:
\begin{equation*}
	\vect{p}_x = |\vect{p}| \sin(\theta_{{\rm p}_z}),
\quad
	\vect{p}_z = |\vect{p}| \cos(\theta_{{\rm p}_z}),
\end{equation*}
\begin{equation*}
	\vect{q}_x = |\vect{q}| \sin(\theta_{\rm q}) \cos(\phi_{\rm q}),
\quad
	\vect{q}_z = |\vect{q}| \cos(\theta_{\rm q}),
\end{equation*}
and the cosine of the angle between momenta~$\vect{p}$ and~$\vect{V}$  is:
\begin{equation*} 
    \cos(\theta_{{\rm p}_z}) 
=
    -\left( |\vect{p}|^2 + |\vect{V}|^2 - |\vect{k}|^2 \right) / \left(2 |\vect{p}| |\vect{V}|\right).
\end{equation*}

Next, we obtain an expression for the double differential cross section in terms of the mediator parameters.
Using the delta functions~$\delta^{(3)}(\vect{p} + \vect{q} - \vect{p'} - \vect{k})$ and~$\delta(E_{p} + q_0 - E_{p'} - E_{k})$, integration is performed over~$\vect{p'}$ and~$\cos(\theta_{\rm q})$ where it is taking into account the expression~\cite{Davoudiasl:2021mjy}:
\begin{multline}
    \delta(E_{p} + q_0 - E_{p'} - E_{k})
= \\ =
    \frac{E_l'}{|\vect{q}| |\vect{V}|}
    \frac{\delta(f(|\vect{q}|) - \cos(\theta_{\rm q}))}{\sqrt{M^2 + |\vect{q}|^2}}
    \Theta(1 - \cos^2(\theta_{\rm q})),
\end{multline}
where $f(|\vect{q}|)$ is the right hand side of the  expression~\eqref{eq:thetaqAsFuncQ} and the Heaviside function~$\Theta(x)$ arises due to the finite limits.
For the invariant volume the following expression reads:
\begin{widetext}
\begin{equation}
	d \Pi 
\to
	\frac{M}{32 \pi^4}
        \Theta(1 - \cos^2(\theta_{\rm q}))
	\frac{dt}{8 M^2}
	\frac{d \phi_{\rm q}}{2 \pi}
	\frac{d\vect{k}}{|\vect{V}|E_l x}
\to
	\frac{2 \pi M}{32 \pi^4}
        \Theta(1 - \cos^2(\theta_{\rm q}))
	\frac{|\vect{k}| E_l}{|\vect{V}|}
	\frac{dt}{8 M^2}
	\frac{d\phi_{\rm q}}{2 \pi}
	d\cos(\theta_{\rm MED}) d x
        \frac{d \phi_{\rm k}}{2 \pi},
\end{equation}
\end{widetext}
where it is taken into account that 
$$dt~=~M(M^2~+~|\vect{q}|^2)^{-1/2} d|\vect{q}|^2.$$

As a result, the double differential cross section for the process of mediator radiation on the nucleus in terms of  $x$ and~$\cos(\theta_{\rm MED})$ takes the form:~\cite{Liu:2016mqv}:
\begin{multline}
	\frac{d \sigma_{2\to3}}{d x d\cos(\theta_{\rm MED})}
= 
	\frac{1}{64 \pi^3}
	\frac{|\vect{k}| E_l}{|\vect{p}| |\vect{k} - \vect{p}|}
\cdot \\ \cdot	
        \int\limits_{t_{\rm min}}^{t_{\rm max}}
	dt
	\frac{1}{8 M^2}
	\int\limits_{0}^{ 2 \pi}
	\frac{d \phi_{\rm q}}{2 \pi}
	\left|\mathcal{M}_{2\to3}^{\rm MED} \right|^2
	%\frac{(c^{\rm MED}_{ll})^2 \alpha^2 }{4 \pi}	
	%\frac{F^2(t)}{t^2}
	%\left|\mathcal{A}_{2\to3}^{\rm MED} \right|^2
	.
\end{multline}
The limits of integration over the photon virtuality are expressed from the condition~$|\cos(\theta_{\rm q})|~<~1$ and takes the values~$t_{\rm max}~=~t(Q_+)$ and~$t_{\rm min}~=~t(Q_-)$ with the introduced notation:
\begin{equation}\label{eq:LimitstVirtualBremssETL}
	Q_{\pm} 
=
    \left|
	\frac{		|\vect{V}| \left[\tilde{u} + 2 M d_{V_0} \right] 
			\pm d_{V_0} \sqrt{ D_0  } 
		 }
		 {2 d_{V_0}^2 - 2 |\vect{V}|^2 }
   \right|,
\end{equation}
where ~$D_0~=~4 M^2 |\vect{V}|^2 + \tilde{u}^2  + 4 M d_{V_0} \tilde{u}$. Note that the condition~$D_0~>~0$ yields an additional kinematic constraint.

It is also worth noting that the given set of variables allows us to obtain a differential 
cross section in terms of outgoing lepton parameters. In particular, one can use the replacement: 
\begin{equation}
	k~\leftrightarrow~p',
\quad
x~\to~y~=~p_0'/p_0,
\quad
	\theta_{\rm MED}~\to~\psi_{l'}.
\end{equation}
where~$\psi_{l'}$ is angle between the direction of the momenta~$\vect{p}$ and~$\vect{p}'$.

The  the matrix element squared for the process~$2\to3$ can be written as:
\begin{equation}
	\overline{\left| \mathcal{M}_{2\to3}^{\rm MED} \right|^2}
	=
	C_{\mathcal{M}}^2 
	\left|\mathcal{A}_{2\to3}^{\rm MED} \right|^2,
\end{equation}
where we define $C_{\mathcal{M}} = c^{\rm MED}_{ll} e^2 Z^2 F_{\rm s}(-q^2)/q^2$.
For the process of radiation of scalar, pseudoscalar, vector and axialvector mediators on the nucleus, the squared amplitudes  take the form, respectively~\cite{Liu:2016mqv,Liu:2017htz,Kirpichnikov:2021jev,Sieber:2023nkq}:
\begin{widetext}
\begin{multline}\label{eq:Ampl2to3ScalarMED}
	\left|\mathcal{A}_{l^{-} N \rightarrow l^{-}N \phi }^{\phi} \right|^2
=
		\frac{\left(\tilde{u} + \tilde{s} \right)^2}{\tilde{u} \tilde{s}} P^{2}
	-   \frac{4 t (k, P)^2 }{\tilde{u} \tilde{s}}
	- 
	 	(m_{\rm MED}^2 - 4 m_{l}^2)
		\frac{\left(\tilde{u} + \tilde{s} \right)^2}{\tilde{u}^2 \tilde{s}^2}
		\left(
		-   P^2 t
		+ 	4 \left(
				\frac{\tilde{u} (p, P) + \tilde{s} (p', P) }{\tilde{u} + \tilde{s}}
			  \right)^2
	\right),
\end{multline}
\begin{multline}\label{eq:Ampl2to3PseudoScalarMED}
	\left|\mathcal{A}_{l^{-} N \rightarrow l^{-}N P }^{\rm P} \right|^2
	=
		\frac{\left(\tilde{u} + \tilde{s} \right)^2}{\tilde{u} \tilde{s}} P^{2}
	-   \frac{4 t (k, P)^2 }{\tilde{u} \tilde{s}}
    -
	 	m_{\rm MED}^2
		\frac{\left(\tilde{u} + \tilde{s} \right)^2}{\tilde{u}^2 \tilde{s}^2}
		\left(
		-   P^2 t
		+ 	4 \left(
			  	\frac{\tilde{u} (p, P) + \tilde{s} (p', P) }{\tilde{u} + \tilde{s}}
			  \right)^2
		\right),
\end{multline}
\begin{multline}\label{eq:Ampl2to3VectorMED}
	\left|\mathcal{A}_{l^{-} N \rightarrow l^{-}N V }^{\rm V} \right|^2
	=
		2 \frac{\tilde{u}^2 + \tilde{s}^2}{\tilde{u} \tilde{s}} P^{2}
	-   \frac{8 t }{\tilde{u} \tilde{s}}
		\left(
		  (p, P)^2 + (p', P)^2 
		+ \frac{2 m_{\rm MED}^2 - \tilde{s} - \tilde{u} - t }{2} P^{2} 
		\right)
	- \\ -	
		2 (m_{\rm MED}^2 + 2 m_{l}^2)
		\frac{\left(\tilde{u} + \tilde{s} \right)^2}{\tilde{u}^2 \tilde{s}^2}
		\left(
		-   P^2 t
		+ 4 \left(
				\frac{\tilde{u} (p, P) + \tilde{s} (p', P) }{\tilde{u} + \tilde{s}}
			\right)^2
		\right),
\end{multline}
\begin{multline}\label{eq:Ampl2to3AxialVectorMED}
	\left|\mathcal{A}_{l^{-} N \rightarrow l^{-}N A }^{\rm A} \right|^2
	=
	2 \frac{\tilde{u}^2 + \tilde{s}^2}{\tilde{u} \tilde{s}} P^{2}
	-   \frac{8 t }{\tilde{u} \tilde{s}}
		\left(
		(p, P)^2 + (p', P)^2 
		- \frac{  \tilde{s} + \tilde{u} + t }{2} P^{2} 
		\right)
	+	4 m_{l}^2
		\frac{\left(\tilde{u} + \tilde{s} \right)^2 P^{2}}
			 {m_{\rm 	MED}^2 \tilde{s} \tilde{u} }
	- \\
	-	\frac{16 m_{l}^2 t (k, P)^2 }
			 {m_{\rm 	MED}^2 \tilde{s} \tilde{u} }	
	-	2 (m_{\rm MED}^2 - 4 m_{l}^2)
		\frac{\left(\tilde{u} - \tilde{s} \right)^2}{\tilde{u}^2 \tilde{s}^2}
		\left(
		-   P^2 t
		+ 	4 \left(
				\frac{\tilde{u} (p, P) + \tilde{s} (p', P) }{\tilde{u} -\tilde{s}}
			\right)^2
		\right),
\end{multline}
where dot products are expressed as:
\begin{equation}
	(P, P) = 4M^2 + t,
\quad
	(k, P) = (p, P) - (p', P),
\quad
	(p, P) 
	%= 2 (p, P_i) - (p, q) 
	= 2 ME_p - (\tilde{s} + t)/2,
	\quad
	( p', P) 
	%= 2 (p', P_i) - (p', q) 
	= 2 M (E_p - E_k) + (\tilde{u} - t)/2.
\end{equation}
\end{widetext}

\section{Analytical expressions for the WW approximation\label{sec:AnaluticalIntegralSection}}

The auxiliary  integrals in this section can be used for analytical integration over an radiated particle angle for the differential cross section in the Weizsäcker-Williams approximation in the case of Tsai-Schiff’s elastic form-factor.

In general, the differential cross section for the compton-like production of considering mediators depends on the angle via the Mandelstam variables as:
\begin{equation}\label{eq:ComptLikeGenDepVarspt}
\frac{|\mathcal{M}^{\rm MED}_{2 \to 2}|^2}{8 \pi (s_2-m_l^2)^2}
=
\frac{P(s_2,t_2,u_2)}{ (s_2-m_l^2)^3(u_2-m_l^2) },
\end{equation}
where~$P(s_2,t_2,u_2)$~-~polynomial function in all arguments e.g.~\eqref{eq:Ampl2to2ScalarMED}~-~\eqref{eq:Ampl2to2AxialVectorMED}.
Also, the  analytically  integrated virtual photon flux with Tsai-Schiff's form-factor reads  as follows~\cite{Kirpichnikov:2021jev}:
\begin{align}
&  \chi_{_{TS}} 
=    \frac{Z^2 t_d^2}{(t_a - t_d)^3}
 %   	\left\lbrace
 \Bigl(   	    \left[
    	        C^{\chi}_{1}
    	    + 
    	        C^{\chi}_{2} t_{\rm min}
    	    \right]
    	+
\nn \\
&
	+
    	    \left[
    	            C^{\chi}_{3}
    	        +   C^{\chi}_{4} t_{\rm min} 
    	    \right]
    	    \ln{\left[ \frac{t_{\rm min} + t_d}{t_{\rm min} + t_a}\right] }
    	    \Bigr) 
%    	\right\rbrace,
    	\label{TsaiFF}
\end{align}
%\begin{equation}\label{TsaiFF}
% \chi_{_{TS}} 
%=    \frac{Z^2 t_d^2}{(t_a - t_d)^3}
%    	\left\lbrace
%    	    \left[
%    	        C^{\chi}_{1}
%    	    + 
%    	        C^{\chi}_{2} t_{min}
 %   	    \right]
 %   	+
 %   	    \left[
 %   	            C^{\chi}_{3}
 %   	        +   C^{\chi}_{4} t_{min} 
 %   	    \right]
 %   	    \ln{\left[ \frac{t_{min} + t_d}{t_{min} + t_a}\right] }
 %   	\right\rbrace,
%\end{equation}
where the functions $C_1^\chi$, $ C_2^\chi$,  $C_3^\chi$ and  $C_4^\chi$ are defined by the following  expressions respectively:
\begin{align*}
&    		C^{\chi}_{1} 
	=
        \Bigl(
		 	\frac{t_d(t_a - t_d)}{t_d + t_{\rm max}}
		+	\frac{t_a(t_a - t_d)}{t_a + t_{\rm max}}
		- 	
 \\   
 &   
 	- 	
		    2 (t_a - t_d)
		+	(t_a + t_d) 
		\ln{\left[ \frac{(t_d + t_{\rm max})}{(t_a + t_{\rm max})}\right] }
		\Bigr),
\end{align*}
\begin{equation*}
		C^{\chi}_{2}
	=
		\left( 
		\frac{t_a - t_d}{t_d + t_{\rm max}}
		+	\frac{t_a - t_d}{t_a + t_{\rm max}}
		+	2 \ln{\left[ \frac{t_d + t_{\rm max}}{t_a + t_{\rm max}}\right] }
		\right),     
\end{equation*}
\begin{equation*}
    				C^{\chi}_{3}
	=
	-	(t_a + t_d),
	\quad
		C^{\chi}_{4}
	=
	-	2.
\end{equation*}

Taking into account the photon flux with Tsai-Schiff's form-factor~\eqref{TsaiFF} and using 
expressions~\eqref{eq_u2_U}~-~\eqref{eq_s2_U}, one can show that the double differential cross 
section  includes the typical terms~$U^l$ and~$U^l \log( U^2 + a(x) )$.  
In particular, one can replace the angular dependence  $\theta_{\rm MED}$ by a $U$-like variable 
(see Eq.~(\ref{eq_vect_U}))  and  represent the analytical integration of the double differential 
cross section over $U$ for the auxiliary integrals~$I_1(x, U, l)~=~U^l$ and $I_2(x, U, l)$.
%Also, coefficient in sums of auxiliary integrals $I_k(x, U, l)$ depends on $x$ and are specific for each type of mediators, $l$ and $k$. 
The integral~$I_2(x, U, l)$ takes the following form:
\begin{equation*}
    I_2(x, U, l )
=
    \int
    U^{l}
    \ln{\left[ \frac{ U^2 + b(x)}
    	            {U^2 + a(x)} \right] }
   dU, \qquad l \in \mathbb{Z},
\end{equation*} 
that allows us to integrate over the angle the differential cross section in the case of scalar, pseudoscalar, vector and axialvector mediators.
For  $l = -1$ and $l = -2$  the integral~$I_2(x, U, l )$ reads, respectively:
\begin{multline*}
		\int \frac{1}{x}\ln{\left( \frac{x^{2} + b}{x^{2} + a}\right) }dx
	=
		 \frac{1}{2}\ln\left(\frac{b}{a}\right) \ln (x^2) 
   - \\ -
   \frac{1}{2} \left[{\rm Li}_2\left( -x^2/b\right) - {\rm Li}_2\left( -x^2/a\right) \right]
\end{multline*}	
\begin{multline*}
		\int \frac{1}{x^{2}}\ln{\left( \frac{x^{2} + b}{x^{2} + a}\right) }dx
	= 
	- \frac{1}{x} \ln{\left( \frac{x^{2} + b}{x^{2} + a}\right)} 
    + \\ +
    2 \left( 
                \frac{\arctan \left( x/\sqrt{b} \right)}{\sqrt{b}} 
            -   \frac{\arctan \left( x/\sqrt{a} \right)}{\sqrt{a}}
        \right) 
\end{multline*}
For $0 \leq l$ the integral~$I_2(x, U, l )$ reads:
\begin{multline*}
		\int x^n \ln{\left( \frac{x^{2} + b}{x^{2} + a}\right) }dx
	=
		\frac{x^{n+1}}{n+1} 
		\ln{\left( \frac{x^{2} + b}{x^{2} + a}\right) } 
	- \\ -
        \frac{2}{n+1} 
        \sum\limits_{m = 0}^{ \frac{n - (n \;{\rm mod}\; 2) }{2} }
        (-1)^{m}
        \frac{x^{n + 1 - 2 m}}{n + 1 - 2 m}
        (b^m - a^m)
    + \\ + 
        (-1)^{\frac{n - (n \;{\rm mod}\; 2) }{2}}
        \frac{2}{n+1}
        R_{21}(x, n, a, b),
\end{multline*}
where  auxiliary  expression $R_{21}(x, n, a, b)$ takes the form:
\[
	R_{21}(x, n, a, b)
=
	\begin{cases}
	       f_{\rm ln}(b) - f_{\rm ln}(a), & n \;{\rm mod}\; 2 = 1, 
        \\
	      f_{\rm atg}(b) - f_{\rm atg}(a), & n \;{\rm mod}\; 2 = 0.
	\end{cases}
\]
with introduced definitions $f_{\rm ln}(b) = (1/2) b^{\frac{n+1}{2}}\ln (x^2 + b)$ and $f_{\rm atg}(b) = b^{\frac{n+1}{2}} \mbox{arctg}(x / \sqrt{b})$.
For values $l < -2$  the integral $I_2(x, U, l)$ takes the following form:
\begin{multline*}
		\int
		x^{n}
		\ln{\left( \frac{x^{2} + b}{x^{2} + a}\right) }dx
	= \\ =
		\frac{x^{n + 1}}
		     {(n + 1) }
		\ln{\left( \frac{x^{2} + b}{x^{2} + a}\right)} 
	+
		R_{22}(x, -n, a, b)
\end{multline*}
where the auxiliary  expression $R_{22}(x, n, a, b)$ for odd values of $n$ reads as follows:
\begin{widetext}
	\begin{multline*}
	R_{22}(x, n, a, b) = 
		\frac{(-1)^{\frac{n - 1}{2}} }{n - 1}
		\left\lbrace  
			\left( \frac{1}{b} \right)^{\frac{n - 1}{2}}
			\ln \left( \frac{x^{2} + b}{x^2}\right)
		-
			\left( \frac{1}{a} \right)^{\frac{n - 1}{2}}
			\ln \left( \frac{x^{2} + a}{x^2}\right)  
		\right\rbrace  
	+ \\ + 
		\frac{2}{n - 1}
		\sum\limits_{k=0}^{\frac{n-5}{2}}
		\frac{(-1)^{k}}{3 + 2 k - n}
		\left( 
			\left( \frac{1}{b} \right)^{k + 1} 
		-
			\left( \frac{1}{a} \right)^{k + 1} 
		\right) 
		\frac{1} 
		{x^{n- 3 - 2k}}
\end{multline*}
and for even values of $n$ is:
\begin{multline*}
    R_{22}(x, n, a, b)  
= 
		(-1)^{\frac{n - 2}{2}}
		\frac{2 }{n - 1}
		\left\lbrace  
			\left( \frac{1}{b} \right)^{\frac{n - 1}{2}}
			\arctan \left(\frac{x}{\sqrt{b}} \right)
		- 
			\left( \frac{1}{a} \right)^{\frac{n - 1}{2} }
			\arctan \left(\frac{x}{\sqrt{a}} \right)  
		\right\rbrace  
	+ \\ + 
		\frac{2}{n - 1}
		\sum\limits_{k=0}^{\frac{n - 4}{2} }
		\frac{(-1)^{k}}{3 + 2 k - n}
		\left( 
			\left( \frac{1}{b} \right)^{k + 1} 
		-
			\left( \frac{1}{a} \right)^{k + 1} 
		\right) 
		\frac{1} 
		{x^{n- 3 - 2k}}
\end{multline*}

The squared  amplitudes of the compton-like processes in the cases of scalar, pseudoscalar, vector and axial-vector mediators, respectively~\cite{Liu:2016mqv,Liu:2017htz,Kirpichnikov:2021jev,Sieber:2023nkq}:
\begin{equation}\label{eq:Ampl2to2ScalarMED}
	\left|\mathcal{A}_{l^{-} \gamma \rightarrow l^{-} \phi }^{\phi} \right|^2
=
	-	\frac{(\tilde{s}_2 + \tilde{u}_2)^2}{\tilde{s}_2\tilde{u}_2}
	+	2 \frac{(m_{\rm MED}^2 - 4 m_l^2)}{\tilde{s}_2\tilde{u}_2}
		\left( 
			\frac{(\tilde{s}_2 + \tilde{u}_2)^2}{\tilde{s}_2\tilde{u}_2}m_l^2
		- \tilde{t}_2 - m_{\rm MED}^2	 
		\right),
\end{equation}
\begin{equation}\label{eq:Ampl2to2PseudoScalarMED}
	\left|\mathcal{A}_{l^{-} \gamma \rightarrow l^{-} P }^{\rm P} \right|^2
	=
	-	\frac{(\tilde{s}_2 + \tilde{u}_2)^2}{\tilde{s}_2\tilde{u}_2}
	+	2 \frac{m_{\rm MED}^2}{\tilde{s}_2\tilde{u}_2}
	\left( 
	\frac{(\tilde{s}_2 + \tilde{u}_2)^2}{\tilde{s}_2\tilde{u}_2} m_l^2
	- \tilde{t}_2 - m_{\rm MED}^2	 
	\right),
\end{equation}
\begin{equation}\label{eq:Ampl2to2VectorMED}
	\left|\mathcal{A}_{l^{-} \gamma \rightarrow l^{-} V }^{\rm V} \right|^2
	=
		4
	-	2 \frac{(\tilde{s}_2 + \tilde{u}_2)^2}{\tilde{s}_2\tilde{u}_2}
	+	4 \frac{(m_{\rm MED}^2 + 2 m_l^2)}{\tilde{s}_2\tilde{u}_2}
	\left( 
	\frac{(\tilde{s}_2 + \tilde{u}_2)^2}{\tilde{s}_2\tilde{u}_2}m_l^2
	- \tilde{t}_2 - m_{\rm MED}^2	 
	\right),
\end{equation}
\begin{equation}\label{eq:Ampl2to2AxialVectorMED}
	\left|\mathcal{A}_{l^{-} \gamma \rightarrow l^{-} A }^{\rm A} \right|^2
	=
	4
	-	\left( 2 + \frac{4 m_l^2}{m_{\rm MED}^2} \right)
		\frac{(\tilde{s}_2 + \tilde{u}_2)^2}{\tilde{s}_2\tilde{u}_2}
	+	4 \frac{(m_{\rm MED}^2 - 4 m_l^2)}{\tilde{s}_2\tilde{u}_2}
	\left( 
	\frac{(\tilde{s}_2 + \tilde{u}_2)^2}{\tilde{s}_2\tilde{u}_2}m_l^2
	- \tilde{t}_2 - m_{\rm MED}^2	 
	\right).
\end{equation}
where we define:
\begin{equation*}
    \tilde{s}_2 = s_2 - m_l^2,
\quad
    \tilde{u}_2 = u_2 - m_l^2,
\quad
    \tilde{t}_2 = t_2 - m_{\rm MED}^2.
\end{equation*}
\end{widetext}

%\section{Typical parameters of the mediator collinear radiation}

\section{Typical mediator radiation angle for the double differential cross section.\label{sec:TypeAngleMED}} 

\begin{figure*}[ht!]
	\center{\includegraphics[scale=0.99]{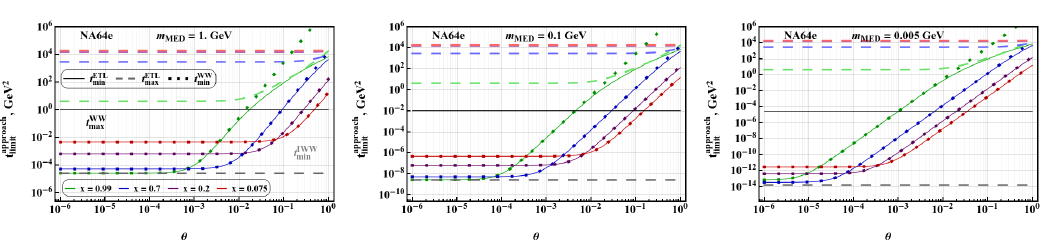}}
	\caption{  The  typical dependence of the maximun, $t_{\rm max}$, and minimum, $t_{ \rm min}$,  momentum squared on the mediator emission angle $\theta_{\rm MED}$ for various approaches 
    (ETL, WW and IWW) and the set of benchmark energy fraction values $x$. Solid (dashed) green, blue, 
    violet and red lines show $t^{\rm ETL}_{\rm min}$ ($t^{\rm ETL}_{\rm max}$) for $x = 0.99$, 
    $x = 0.7$, $x=0.2$, and $ x= 0.075$, respectively. The same color dotted lines
    depict $t^{ \rm WW}_{\rm min}$ for given benchmark set of $x$. The typical magnitude of 
    $t^{\rm WW}_{\rm max}$ is estimated to be $ t^{\rm WW}_{\rm max} \simeq m^2_{\rm MED} + m_l^2$ 
    according to Refs.~\cite{Liu:2016mqv,Liu:2017htz}. The latter doesn't depend on both $x$ and 
    $\theta$ and shown by the solid black line. We note that it implies  
    $t^{\rm WW}_{\rm max} \equiv t^{\rm IWW}_{\rm max}$. Finally, the constant value of $ t^{\rm IWW}_{ \rm min}$ is 
    defined by  Eq.~(\ref{tminIWWdefinition}) and shown by gray dashed line. 
	}\label{fig:tLimits}
\end{figure*}

In Fig.~\ref{fig:DsDxDthetaWWandETLScalarMED}
we show that  the WW differential cross-section falls off much more steeply
with $\theta_{\rm MED}$ than the ETL one. That can be explained by the reduced 
phase space of the WW cross section compared to the ETL one. In Fig.~\ref{fig:tLimits}
we show the typical $t_{\rm max}$ and $t_{\rm min}$ as a functions of $\theta_{\rm MED}$ for the  
approaches of the  interest (ETL, WW, and IWW). The intersection points of $ t_{\rm max}^{\rm WW}$ and $ t_{\rm min}^{\rm WW}$ functions imply $ t_{\rm max}^{\rm WW} =  t_{\rm min}^{\rm WW}$, therefore  these points  lead to the 
shrinking of the WW phase space to zero (the double differential cross section falls 
rapidly to zero at the point $\theta^{\rm 2D}_{\rm cut}$).
In particular, the  cut-off angle is estimated to be 
\begin{multline}
    \theta^{\rm 2D}_{\rm cut}
\simeq
    \frac{1}{\sqrt{E_l^2x}}
    \left( 
        2 E_l \sqrt{m_{\rm MED}^2 + m_l^2} (1 - x)
    - \right. \\ \left. -
        m_{\rm MED}^2 \frac{(1 - x)}{x}
    -   m_l^2 x 
    \right)^{1/2},
\end{multline}
from $ t_{\rm max}^{\rm WW} =  t_{\rm min}^{\rm WW}$ equation. 
For instance,  in Fig.~\ref{fig:tLimits} for $x=0.99$  and  $m_{\rm MED} = 1~\mbox{GeV}$ we show that $ t_{\rm max}^{\rm WW} =  t_{\rm min}^{\rm WW}$ at $\theta_{\rm MED} \simeq 10^{-2}$ for NA64e experiment. 
Thus, the corresponding WW differential cross section cuts  at $\theta_{\rm MED} \simeq 10^{-2}$, that is shown in Fig.~\ref{fig:DsDxDthetaWWandETLScalarMED} by  solid green line in the bottom row  
for $m_{\rm MED} = 1~\mbox{GeV}$.

On the other hand, for the ETL method there is intersection
 $ t_{\rm min}^{\rm ETL } = t_{\rm max}^{\rm ETL }$ for the  sufficiently large angles 
$\theta_{\rm MED} \gtrsim 1$ and heavy mediator $m_{\rm MED} \simeq 1~\mbox{GeV}$. 
This  means  that  the ETL cross section dependence on $\theta_{\rm MED}$ is more flatter in this 
region.

Let us estimate  a 
typical mediator radiation angle  in the analytical form for the double differential cross section.
We consider the IWW approximation that implies 
Eq.~(\ref{tminIWWdefinition}), as a result,  the virtual photon 
flux (\ref{ChiDefininition1}) does not depend on $x$ and 
$\theta_{\rm  MED}$. The simplified IWW approach allows us to estimate
qualitatively  $\theta_{\rm MED}$  associated 
with collinear  emission of the mediators.

One can write mediator radiation angle dependence of double-differential cross section as:
\begin{multline}
    	\left.\frac{d \sigma ( p + P_i \rightarrow  p' + P_f + k )}{dx d\theta_{\rm MED}}\right|_{\rm IWW} 
= \\ =
        C_{\sigma_{2\to3}^{\rm IWW}}(x)
        \frac{\sin(\theta_{\rm MED})
            \left|\mathcal{A}_{l^{-} \gamma \rightarrow l^{-} {\rm MED} }^{{\rm MED}} \right|^2}
             {U^2(\theta_{\rm MED})}, \label{DoubleDiffCSIWW}
\end{multline}
where $U~>~0$ variable is defined by Eq.~\eqref{eq_vect_U} 
\[
    C_{\sigma_{2\to3}^{\rm IWW}}(x)
=
        \frac{4 \pi (c^{\rm MED}_{ll})^2 \alpha}{8\pi}
       \frac{\alpha \chi}{ \pi }
        E_l^2(1 - x)
        \sqrt{x^2 - \frac{m_{\rm MED}^2}{(E_l)^2}},
\]
and the squared amplitude for Compton-like process is~\cite{Sieber:2023nkq}:
\[
    \left|\mathcal{A}_{l^{-} \gamma \rightarrow l^{-} {\rm MED} }^{{\rm MED}} \right|^2
=
    C_1^{\rm MED}(x)
-   \frac{C_2^{\rm MED}(x)}{U(\theta_{\rm MED})}
+   \frac{C_3^{\rm MED}(x)}{U^2(\theta_{\rm MED})}.
\]
 The explicit forms of the coefficients~$C_i^{\rm MED}$ for scalar,~S, vector,~V, pseudoscalar,~P, and axialvector,~A, types of mediator read~\cite{Kirpichnikov:2021jev,Sieber:2023nkq}:
\begin{equation*}
    C^{S}_{1} = x^2 / (1 - x),
\quad
    C^{S}_{2}  = 2 \left( m_{\rm MED}^2 - 4 m_{l}^2\right) x,
\end{equation*}
\begin{equation*}
    C^{S}_{3}   
= 
    2 \left( m_{\rm MED}^2 - 4 m_{l}^2\right) 
    \left( m_{\rm MED}^2 (1 - x) + m_{l}^2 x^2 \right),
\end{equation*}
\begin{equation*}
    C^{V}_{1}  
= 
	2 \frac{(2 - 2 x + x^2)}{1 - x}, 
\quad
	C^{V}_{2}  
=	
	4 \left( m_{\rm MED}^2 + 2 m_{l}^2\right) x,
\end{equation*}
\begin{equation*}
    C^{V}_{3}  
=
	4 \left( m_{\rm MED}^2 + 2 m_{l}^2\right)
	\left( m_{\rm MED}^2 (1 - x) + m_{l}^2 x^2 \right),
\end{equation*}
\begin{equation*}
    C^{P}_{1}  =  x^2 / (1 - x),
\quad
    C^{P}_{2}   =	 2 m_{\rm MED}^2 x,
\end{equation*}
\begin{equation*}
    C^{P}_{3} = 2 m_{\rm MED}^2 \left( m_{\rm MED}^2 (1 - x) + m_{l}^2 x^2 \right),
\end{equation*}
\begin{equation*}
    C^{A}_{1} 
= 
    4 +  \left( 2 + \frac{4 m_l^2}{m_{\rm MED}^2} \right) \frac{x^2}{1-x},
\;\;
    C^{A}_{2} = 4 (m_{\rm MED}^2 - 4 m_l^2) x,
\end{equation*}
\begin{equation*}
    C^{A}_{3} 
= 
    4 (m_{\rm MED}^2 - 4 m_l^2) 
    \left( m_{\rm MED}^2 (1 - x) + m_{l}^2 x^2 \right).
\end{equation*}
%where we denoted~$R~=~\left( m_{\rm MED}^2 (1 - x) + m_{l}^2 x^2 \right)$.

For the collinear singularity region $x\simeq 1$  the dominant contribution to the double-
differential cross section is provided  by the terms  
$$\sin (\theta_{\rm MED}) C^{\rm MED}_1(x) /
U^2(\theta_{\rm MED})$$ 
for all mediators, since 
$C^{\rm MED}_1(x) \propto 1/(1-x)$ in the collinear emission approach $(1-x) \ll 1$.
So that, one  can find typical $\theta_{\rm MED}$ that maximizes the Eq.~(\ref{DoubleDiffCSIWW}). This condition implies that 
\begin{equation}
    \tan(\theta_{\rm typ}^{\rm 2D}) 
= 
   (1/2) U / (U)'_{\theta_{\rm MED}},
   \label{LinkTan}
\end{equation}
where $(U)'_{\theta_{\rm MED}}$ is a derivation of (\ref{eq_vect_U}) with respect to 
$\theta_{\rm MED}$.
Finally, implying a  small values~$\tan(\theta_{\rm typ}^{\rm 2D})~\simeq~\theta_{\rm typ}^{\rm 2D} \ll 1$, one can resolve aon Eq.~(\ref{LinkTan}) and obtain the  typical mediator radiation angle in the following form:
\begin{equation}\label{eq:TypeAngleMEDfor2D}
    \theta_{\rm typ}^{\rm 2D}
=
    \sqrt{\frac{1}{ 3 E_l^2 x}
          \left(\frac{m_{\rm MED}^2(1-x)}{x} +   m_l^2 x\right)}.
\end{equation}
In Fig.~\ref{fig:DsDxDthetaWWandETLScalarMED} we depict $\theta_{\rm typ}^{\rm 2D}$ by black dots for various $x$ and set of masses $m_{\rm MED}=(5~\mbox{MeV},  1~\mbox{GeV})$. 
Remarkably that Eq.~(\ref{eq:TypeAngleMEDfor2D}) is a fairly good estimate also for the 
typical angles  of  both ETL and WW  cross sections.

Moreover, by substituting the expression for~$x_{\rm min}$ and~$x_{\rm max}$ into Eq.~(\ref{eq:TypeAngleMEDfor2D}), one can get  the bounds on the typical mediator radiation angle:
\begin{equation*}
    \frac{m_l}{3 E_l} \left(\frac{m_l}{E_l} + \frac{m_{\rm MED}^2}{E_l^2}\right)
    %\frac{m_l \left(m_l + \frac{E_l m_{\rm MED}^2}{(E_l - m_l)^2}\right)}{3 E_l^2}
\lesssim
    \left( \theta_{\rm typ}^{\rm 2D} \right)^2
< 
    \frac{E_l(E_l - m_{\rm MED}) + m_l^2 }{3 E_l^2}.
\end{equation*}
Taking into account the expression above in the case of a muon beam in the region of light mediator masses,~$m_{\rm MED}~\ll~m_l$, we get~$\frac{m_l^2}{3 E_l^2}~\lesssim~\left( \theta_{\rm typ}^{\rm 2D} \right)^2$.
Indeed, one can see from Fig.~\ref{fig:DsDxDthetaWWandETLScalarMED} that for muon beam at~$m_{\rm MED}~=~0.005~\mbox{GeV}$ the typical mediator radiation angles is~$\theta_{\rm typ}^{\rm 2D}~\simeq~m_l/\left(\sqrt{3} E_l\right)$. That implies  $\theta_{\rm typ}^{\rm 2D}\simeq 3.8\times~10^{-4}$ and~$\theta_{\rm typ}^{\rm 2D} \simeq 4.0\times~10^{-3}$ for NA64$\mu$ and M$^3$ experiments, respectively.
It is also worth noting that both the cut-off and the typical mediator radiation angles of the double differential cross section  scale with the energy as~$\theta^{\rm 2 D}~\propto~1/E_l$.

\bibliography{bibl}

\end{document}